\documentclass[aap,MSNbibl,seceqn,dvips]{arximspdf}

\doi{10.1214/12-AAP872} 
\volume{23}
\issue{4}
\pubyear{2013}
\firstpage{1355}
\lastpage{1376}

\makeatletter
\newcommand{\rrvert}{\vert}
\newcommand{\llvert}{\vert}

\newtheorem{theorem}{Theorem}[section]
\newtheorem{lem}[theorem]{Lemma}
\newtheorem{prop}[theorem]{Proposition}

\newproclaim{defn}[theorem]{Definition}
\newproclaim{rem}[theorem]{Remark}
\newproclaim{conv}[theorem]{Convention}

\newcommand{\Real}{\mathbb R}
\newcommand{\Natural}{\mathbb N}
\newcommand{\such}{\mid}
\newcommand{\nin}{{n \in\Natural}}
\newcommand{\kin}{{k \in\Natural}}

\newcommand{\Pre}{\mathcal{P}}
\newcommand{\prob}{\mathbb{P}}
\newcommand{\oprob}{\mathbb{P}}

\newcommand{\lzp}{\mathbb{L}^0_+}

\newcommand{\up}{\mathsf{u}\prob}
\newcommand{\qprob}{\mathbb{Q}}
\newcommand{\expec}{\mathbb{E}}
\newcommand{\expecp}{\expec}
\newcommand{\expecq}{\expec_\qprob}
\newcommand{\var}{\operatorname{\mathsf{Var}}}
\newcommand{\F}{\mathcal{F}}
\newcommand{\G}{\mathcal{G}}
\newcommand{\ud}{\,\mathrm d}

\newcommand{\Time}{\mathbb{T}}

\newcommand{\X}{\mathcal{X}}

\newcommand{\Y}{\mathcal{Y}}

\newcommand{\oxs}{\overline{\X}}

\newcommand{\Sem}{\mathcal{S}}

\newcommand{\tX}{\widetilde{X}}
\newcommand{\hX}{\widehat{X}}
\newcommand{\hY}{\widehat{Y}}

\newcommand{\llbracket}{[\![}
\newcommand{\rrbracket}{]\!]}

\newcommand{\dbra}[1]{\llbracket #1 \rrbracket}
\newcommand{\C}{\mathcal{C}}
\newcommand{\D}{\mathcal{D}}
\newcommand{\bF}{\mathbf{F}}
\newcommand{\indic}{\mathbb{I}}
\newcommand{\dfn}{:=}

\makeatother

\begin{document}
\begin{frontmatter}

\title{On the closure in the Emery topology of semimartingale
wealth-process sets\thanksref{T1}}
\runtitle{Closure in the Emery topology of wealth-process sets}

\thankstext{T1}{Supported in part by the NSF Grant DMS-09-08461. Any opinions,
findings, and
conclusions or recommendations expressed in this material are those of
the author and do not necessarily reflect the views of the National
Science Foundation.}

\begin{aug}
\author[A]{\fnms{Constantinos} \snm{Kardaras}\corref{}\ead[label=e1]{K.Kardaras@lse.ac.uk}}
\runauthor{C. Kardaras}
\affiliation{London School of Economics and Political Science}
\address[A]{Department of Statistics\\
London School of Economics\\
\quad and Political Science\\
10 Houghton st, London\\
WC2A 2AE\\
United Kingdom\\
\printead{e1}} 
\end{aug}

\received{\smonth{8} \syear{2011}}
\revised{\smonth{3} \syear{2012}}

%
\begin{abstract}
A wealth-process set is abstractly defined to consist of nonnegative
c\`adl\`ag processes containing a strictly positive semimartingale and
satisfying an intuitive re-balancing property. Under the condition of
absence of arbitrage of the first kind, it is established that all
wealth processes are semimartingales and that the closure of the
wealth-process set in the Emery topology contains all ``optimal''
wealth processes.
\end{abstract}

%
\begin{keyword}[class=AMS]
\kwd{60H99}
\kwd{60G44}
\kwd{91B28}
\kwd{91B70}
\end{keyword}
\begin{keyword}
\kwd{Wealth-process sets}
\kwd{semimartingales}
\kwd{Emery topology}
\kwd{utility maximization}
\end{keyword}

\end{frontmatter}

\section*{Introduction}\label{intro}

In financial modeling, it is customary to start by describing a set of
wealth processes that can be achieved in some elementary way. Concrete
examples include:
\begin{itemize}
\item wealth processes arising from finite combinations of buy-and-hold
strategies;
\item wealth processes resulting from taking positions on a finite
number of investment assets, when there is an infinite number of such
assets available in the market. This is the case in the theoretical
modeling of bond markets, where there exist zero-coupon bonds with a
continuum of maturities---see, for example,~\cite{BDMKR} and \cite
{MR2187311}. Another case is the approximation of ``large'' financial
markets, as is discussed in~\cite{MR2178505}.
\end{itemize}

Although such initial descriptions of available wealth processes are
natural and unquestionable, the thus-constructed classes are typically
insufficient for analysis. Indeed, important problems like portfolio
optimization and hedging of contingent claims might fail to have
solutions within the class of wealth processes, if the latter is
lacking any reasonable closedness property. Therefore, the need arises
to pass to the closure, in some appropriate sense, of these elementary
wealth-process sets. Such passage is a rather subtle issue: although
the closure should be large enough to ensure that all ``interesting''
(or ``optimal'') elements are there, the need to keep a tight financial
interpretation of the resulting enlarged wealth-process set dictates
that fine topologies are required.

In the literature, a balance between the aforementioned opposing forces
has to be resolved individually for each problem-at-hand. For example,
when wealth processes are defined using simple integrands (i.e., finite
combinations of buy-and-hold strategies) against a finite-dimensional
semimartingale integrator, the class of all stochastic integrals using
general predictable integrands turns out to be the appropriate
enlargement---indeed, this has been demonstrated in a number of papers,
with~\cite{MR1304434,MR1722287} and~\cite{MR2023886} being the
ones related to questions of market viability and optimization that are
close to the spirit of the present discussion. In fact, the class of
stochastic integrals using general predictable integrands coincides
with the closure of the set of all simple integrals in the so-called
\textit{Emery} (or \textit{semimartingale}) \textit{topology}, introduced in
\cite{MR544800}. An enlargement of the initial wealth-process set using
limits of semimartingales in the Emery topology is also utilized in
\cite{MR2178505} and~\cite{MR2187311}, when approximating stochastic
integrals with respect to an infinite-dimensional integrator via
stochastic integrals with integrands having only a finite number of
nonzero coordinates.

The Emery topology is extremely strong and, at the same time, very
natural when dealing with semimartingales. The purpose of this paper is
to show, in an abstract and general setting, that it is the closure of
wealth-process sets in the Emery topology that is indeed appropriate if
one wants to ensure that ``optimal'' elements are contained in the
enlarged class of wealth processes. For the sake of generality,
admissible wealth-process sets are defined in an abstract way, asking
that they consist of nonnegative adapted c\`adl\`ag processes containing
one strictly positive semimartingale (which can be, e.g., the outcome
of investing in a locally riskless asset) and satisfying an intuitive
re-balancing property, called \textit{fork-convexity} in \cite
{MR1883202}. It is first established that, under the mild condition of
absence of arbitrage of the first kind in the market, all wealth
processes are semimartingales---because of this fact, taking the
closure of the wealth-process set in the Emery topology becomes both
relevant and possible. Following this preliminary result, the main
message of the paper is the following: the closure of wealth-process
sets in the Emery topology is rich enough in order to allow for
solutions to expected utility maximization problems. More precisely,
even though an optimal wealth process might not exist in the original
wealth-process set, one can find a sequence of ``nearly-optimal''
wealth processes that has a limit in the Emery topology, and the latter
limit is indeed optimal in the enlarged wealth-process set.\vadjust{\goodbreak}

The results of this paper serve as a guideline in efficiently defining
enlargements of wealth-process sets, after an elementary and acceptable
initial description has been carried out. The fineness of the Emery
topology on semimartingales ensures that the resulting enlarged
wealth-process set will be quite close to the original one. It is
exactly the general and abstract nature of the definition of
wealth-process sets that makes the hereby presented results valuable.
Needless to say, when faced with a specific application one should aim
for more ``intrinsic'' and elegant descriptions of the closure of
elementary wealth-process sets in the Emery topology.

The structure of the paper is simple. Section~\ref{secmain} contains
all the set-up, discussion and results. All proofs are deferred to
Section~\ref{secproofs}. 

\section{Results} \label{secmain}

\subsection{Preliminaries}\label{subsecprelims}

Throughout, $T \in(0, \infty)$ will be denoting a fixed financial
planning horizon. We shall be working on a stochastic basis $(\Omega,
\mathcal{F},\break \mathbf{F}, \prob)$,
where $\bF= (\F_t)_{t \in[0, T]}$ is a filtration satisfying the
usual hypotheses of right-continuity and saturation by $\prob$-null
sets of $\F$. Without loss of generality, we assume that $\F_0$ is
trivial modulo $\prob$ and that $\F= \F_T$. Random variables are
identified modulo $\prob$-a.s. equality. Stochastic processes that are
indistinguishable modulo $\prob$ are also identified. A c\`adl\`ag
(right-continuous with left limits) stochastic process $X$ will be
called \textit{nonnegative} (resp., \textit{strictly positive}) if
$\prob[\inf_{t \in[0, T]} X_t \geq0 ] = 1$ (resp., if
$\prob
[\inf_{t \in[0, T]} X_t > 0 ] = 1$).

The class of semimartingales on $(\Omega, \mathcal{F}, \mathbf{F},
\prob)$ is denoted by $\Sem$. If $X
\in\Sem$ and $\eta$ is a predictable and $X$-integrable process,
$\eta
\cdot X$ denotes the stochastic integral of $\eta$ with respect to
$X$---by convention, $(\eta\cdot X)_0 = \eta_0 X_0$. Let $\Pre_1$ be
the set of predictable processes $\eta$ with $|\eta| \leq1$. For $X
\in\Sem$, define
\[
\lceil X \rceil_{\Sem} \dfn\sup_{\eta\in\Pre_1} \expec\Bigl[1 \wedge
\Bigl(\sup_{t \in[0, T]} \bigl\llvert(\eta\cdot X)_t\bigr\rrvert
\Bigr) \Bigr],
\]
where ``$\expecp$'' is used to denote expectation under $\prob$ and
``$\wedge$'' is used to denote the minimum operation. The metric $\Sem
\times\Sem\ni(X, X') \mapsto\lceil X - X' \rceil_{\Sem}$ induces the
\textit
{Emery topology} on $\Sem$, introduced in~\cite{MR544800}. Whenever
$\lim_{n \to\infty} \lceil X^n - X \rceil_{\Sem} = 0$,
we write $\mathcal{S}\mbox{-}\lim_{n \to\infty}X^n = X$.
Convergence in the
Emery topology is extremely strong; for example, it implies uniform
convergence in probability and (as Proposition \ref
{propstabilityofsem-conv} later in the text shows) convergence of
quadratic variations.

\subsection{Financial set-up}

The first line of business is to model the class of wealth processes
available to an investor with (normalized) unit initial capital. The
wealth-process set will be defined in a rather abstract and generally
encompassing way: any reasonable class of (potentially, constrained)
nonnegative wealth processes resulting from frictionless trading that
has appeared in the literature falls within its scope.

%
\begin{defn} \label{defnwealth-process}
A set $\X$ of stochastic processes will be called a \textit
{wealth-process set} if:
\begin{longlist}[(3)]
\item[(1)] Each $X \in\X$ is a nonnegative c\`adl\`ag process with $X_0 = 1$.
\item[(2)] There exists a strictly positive semimartingale in $\X$.
\item[(3)]$\X$ is \textit{fork-convex}: for any $s \in[0, T]$, $X \in\X$,
any strictly positive processes $X' \in\X$ and $X'' \in\X$, and any
$[0,1]$-valued $\F_s$-measurable random variable $\alpha$, the process
%
\begin{equation}
\label{eqfork-convexity}\quad
[0, T] \ni t \mapsto X_t
\indic_{ \{t < s \}} + \bigl(\alpha\bigl(X_s / X'_s
\bigr) X_t' + (1 - \alpha) \bigl(X_s /
X''_s \bigr) X_t''
\bigr) \indic_{ \{s \leq t \}}
\end{equation}
is also an element of $\X$.
\end{longlist}
\end{defn}
In Definition~\ref{defnwealth-process} of a wealth-process set,
fork-convexity corresponds to the possibility of re-balancing. In fact,
(\ref{eqfork-convexity}) exactly describes the wealth generated when
a financial agent invests according to $X$ up to time $s$, and then
reinvests a fraction $\alpha$ of the money in the wealth process
described by $X'$ and the remaining fraction ($1 - \alpha$) in the
wealth process described by~$X''$. On the other hand, condition (2) is
always true when a locally riskless investment opportunity exists
leading to a wealth process that is adapted, right-continuous and nondecreasing.
%
\begin{defn} \label{defnNA1}
Let $\X$ be a wealth-process set. For $x \in(0, \infty)$, define $\X
(x) \dfn\{x X \such X \in\X\}$. We say that there are
opportunities
for arbitrage of the first kind in the market if there exists an $\F
_T$-measurable random variable $\xi$ such that:
\begin{itemize}
\item$\prob[\xi\geq0] = 1$ and $\prob[\xi> 0] > 0$;
\item for all $x \in(0, \infty)$ there exists $X \in\X(x)$, which
may depend on $x$, with $\prob[X_T \geq\xi] = 1$.
\end{itemize}
If there are \textit{no} opportunities for arbitrage of the first kind,
we shall say that condition $\mathrm{NA}_1$ holds.
\end{defn}

In the context of Definition~\ref{defnNA1}, $\X(x)$ represents all
wealth processes that are available to an investor with initial capital
$x \in(0, \infty)$. Keeping this in mind, the definition of arbitrage
of the first kind is very natural: regardless of how minuscule the
initial capital is, an investor is able to choose a wealth process that
will result in an outcome which dominates $\xi$, the latter being a
nonnegative random variable which is strictly positive on an event of
strictly positive probability.

\subsection{Results} \label{subsecresults}

We are ready to present the findings of the paper; proofs are deferred
to Section~\ref{secproofs}.

We start with a result stating that condition $\mathrm{NA}_1$ already enforces
a semimartingale structure on wealth-process sets. Similar results, in
the case where the wealth-process set is defined as nonnegative \textit
{simple}\vadjust{\goodbreak} stochastic integrals (using linear combinations of
buy-and-hold strategies) against a c\`adl\`ag adapted process have been
established in~\cite{MR1304434}, Section 7,~\cite{MR2832419} and
\cite{BSV}.

\begin{theorem} \label{thmsmarts}
Let $\X$ be a wealth-process set, and assume condition $\mathrm{NA}_1$.
Then, every process in $\X$ is a semimartingale.
\end{theorem}

In view of Theorem~\ref{thmsmarts}, whenever $\X$ is a wealth-process
set such that condition $\mathrm{NA}_1$ is valid, we define $\oxs$ as the
closure of $\X$ in the Emery topology. It follows that $\oxs$ is also a
wealth-process set that is further closed in the Emery topology.
Indeed, the only fact that is not trivial is that $\oxs$ is
fork-convex. Fix $s \in[0, T]$, $X \in\oxs$, any strictly positive
processes $X' \in\oxs$ and $X'' \in\oxs$, and any $[0,1]$-valued
$\F_s$-measurable random variable $\alpha$. Pick $\X$-valued sequences
$(X^n)_{\nin}$, $((X')^n)_{\nin}$ and $((X'')^n)_{\nin}$ such that
$\mathcal{S}\mbox{-}\lim_{n \to\infty}
X^n = X$, $\mathcal{S}\mbox{-}\lim_{n \to\infty}(X')^n = X'$ and
$\mathcal{S}\mbox{-}\lim_{n \to\infty}(X'')^n = X''$. It can be assumed
without loss of generality that the sequences $((X')^n)_{\nin}$ and
$((X'')^n)_{\nin}$ consist of strictly positive wealth processes in
$\X
$; otherwise, with $\chi\in\X$ being strictly positive, one may
replace $(X')^n$ with $(1 - n^{-1})(X')^n+ n^{-1} \chi$ and $(X'')^n$
with $(1 - n^{-1})(X'')^n+ n^{-1} \chi$ for all $\nin$; the previous
are strictly positive wealth processes, and $\mathcal{S}\mbox{-}\lim_{n
\to\infty} ((1 - n^{-1})(X')^n+ n^{-1} \chi) = X'$
as well as $\mathcal{S}\mbox{-}\lim_{n \to\infty} ((1 -
n^{-1})(X'')^n+ n^{-1} \chi) = X''$ still hold. It follows that the
process $\psi^n$, defined via $\psi^n_t \dfn X^n_t \indic_{ \{t
< s \}}
+ (\alpha(X^n_s / (X')^n_s ) (X')^n_t + (1 - \alpha
) (X^n_s / (X'')^n_s ) (X'')^n_t ) \indic_{ \{
s \leq t \}}$ for $t \in[0,
T]$ is an element of $\X$ for all $\nin$. Furthermore, it is
straightforward from the definition of $\Sem$-convergence that the
sequence $(\psi^n)_{\nin}$ converges in the Emery topology to the
process in (\ref{eqfork-convexity}). This establishes the
fork-convexity of $\oxs$.

We proceed in giving justice to the claim (made in the \hyperref[intro]{Introduction})
that $\oxs$ already contains all interesting ``optimal'' elements, by
examining the problem of expected utility maximization. Let $U\dvtx  (0,
\infty) \mapsto\Real$ be a strictly increasing, strictly concave and
continuously differentiable function, satisfying the Inada conditions
$\lim_{x \downarrow0} U'(x) = \infty$ and $\lim_{x \uparrow\infty}
U'(x) = 0$. Also, set $U(0) \dfn\lim_{x \downarrow0} U(x)$ in order
to accommodate possibly zero wealth. With $\X$ being a wealth-process
set such that $1 \in\X$, define the \textit{indirect utility function}
$u\dvtx  (0, \infty) \mapsto\Real\cup\{\infty\}$ via $u(x)
= \sup_{X
\in\X(x)} \expec[ U(X_T) ]$ for $x \in(0, \infty)$. (In
order for an expression of the form $\expec[ U(X_T) ]$, where
$X \in\X(x)$ for some $x \in(0, \infty)$, to be well defined, the
usual convention $\expec[ U(X_T) ] = - \infty$ whenever
$\expec[ 0 \wedge U(X_T) ] = - \infty$ is used. Also, note
that $u \geq U$ follows from $1 \in\X$, which implies that $u(x) > -
\infty$ for all $x \in(0, \infty)$.) In accordance with Definition
\ref
{defnNA1}, set $\oxs(x) \dfn\{x X \such X \in\oxs\}$
for $x \in
(0, \infty)$. It is not a priori clear that $\sup_{X \in\X(x)}
\expecp
[U(X_T) ] = \sup_{X \in\oxs(x)} \expecp
[U(X_T) ]$ holds for $x
\in(0, \infty)$; however, as Theorem~\ref{thmutilmax} states, this
is indeed true under assumption $\mathrm{NA}_1$. What \textit{is} clear
is that,
in general, maximal expected utility will not be achieved by a wealth
process in $\X(x)$ for $x \in(0, \infty)$; as it turns out, maximal
utility \textit{can} be achieved by a process in $\oxs(x)$, at least\vadjust{\goodbreak}
under condition $\mathrm{NA}_1$ and the validity of the following:
\renewcommand{\theequation}{FIN-DUAL}
\begin{equation}\label{eqFIN-DUAL}\quad
\sup_{x > 0} \bigl\{u(x) - x y \bigr
\}
< \infty\qquad\mbox{holds for all } y \in(0, \infty).
\end{equation}
Furthermore, for all $x \in(0, \infty)$, the optimal wealth process in
$\oxs(x)$ along with its expected utility can be approximated
arbitrarily by wealth processes in $\X(x)$. The exact statement follows.
%
\begin{theorem} \label{thmutilmax}
Let $\X$ be a wealth process set with $1 \in\X$, and suppose that
condition $\mathrm{NA}_1$ is valid.
\begin{longlist}[(2)]
\item[(1)]$u(x) \dfn\sup_{X \in\X(x)} \expecp[U(X_T) ] =
\sup_{X
\in
\oxs(x)} \expecp[U(X_T) ]$ holds for all $x \in(0,
\infty)$.
\item[(2)] Suppose that (\ref{eqFIN-DUAL}) is also valid. Then, for all $x
\in(0, \infty)$, there exists $\hX(x) \in\oxs(x)$ satisfying
$\expec
[U(\hX(x)_T) ] = u(x) < \infty$; furthermore, there exists
an $\X
(x)$-valued sequence $ (X^n(x) )_{\nin}$ such that both\break
$\mathcal{S}\mbox{-}\lim_{n \to\infty}X^n(x)
= \hX(x)$ and $\lim_{n \to\infty}\expec[U(X^n(x)_T) ] =
\expec
[U(\hX
(x)_T) ] = u(x)$ hold.
\end{longlist}
\end{theorem}
%
\begin{rem}
For $U\dvtx  (0, \infty) \mapsto\Real$ as before, define $U(\infty) \dfn
\lim_{x \uparrow\infty} U(x)$.

When $U(\infty) = \infty$ and condition $\mathrm{NA}_1$ fails for a
wealth-process set $\X$ with \mbox{$1 \in\X$}, it is straightforward that
$u(x) = \infty$ holds for all $x \in(0, \infty)$. On the other hand,
condition (\ref{eqFIN-DUAL}) always implies that $u$ is
finitely-valued. It then follows that, when $U(\infty) = \infty$ and
$\X
$ is a wealth process with $1 \in\X$, (\ref{eqFIN-DUAL}) is
sufficient to have both statements of Theorem~\ref{thmutilmax}
valid, since condition $\mathrm{NA}_1$ is indirectly forced.

Note also that when $U(\infty) < \infty$ condition (\ref{eqFIN-DUAL})
is always trivially valid; therefore it does not have to be assumed in
statement (2) of Theorem~\ref{thmutilmax}.
\end{rem}
%
\begin{rem}
The proof of the existence of optimal wealth processes in statement (2)
of Theorem~\ref{thmutilmax} heavily depends on the two seminal
papers of Kramkov and Schachermayer~\cite{MR1722287,MR2023886}. At
first sight, the setting of the present paper does not match the one of
\cite{MR1722287} and~\cite{MR2023886}---indeed, in the latter papers
the wealth-process sets are modeled via outcomes of stochastic
integrals with respect to a finite-dimensional semimartingale
integrator. However, \cite{MR1722287}~and~\cite{MR2023886} contain
certain ``abstract results'' that we shall be eventually able to use in
order to show the validity of Theorem~\ref{thmutilmax}.
\end{rem}

In fact, there is an intermediate result used in order to establish
Theorem~\ref{thmutilmax}, which is in some sense more fundamental.
%
\begin{theorem} \label{thmmain}
Let $\X$ be a wealth-process set, and assume condition $\mathrm{NA}_1$.
Then, for any $\qprob\sim\prob$ there exists a strictly positive
$\hX^{\qprob} \in\oxs$ such that $X / \hX^\qprob$ is a $\qprob
$-supermartingale for all $X \in\oxs$.\vadjust{\goodbreak}
\end{theorem}
%
\begin{rem}
Theorem~\ref{thmmain} is related to the idea of \textit{change of
num\'eraire}---see~\cite{MR1381678}. Using notation from Theorem \ref
{thmmain},
the probability $\qprob$ is an equivalent supermartingale measure in
the market where wealth is denominated by $\hX^\qprob\in\oxs$. In
accordance to the terminology of~\cite{Long90,MR1849424} and
\cite{MR2335830}, one can call $\hX^\qprob$ the \textit{num\'eraire
portfolio} in $\X$ under the probability $\qprob$.
\end{rem}
%
\begin{rem}
We elaborate on how Theorems~\ref{thmutilmax} and~\ref{thmmain}
are connected. Technicalities aside, the num\'eraire portfolio $\hX
^\qprob$ in the notation of Theorem~\ref{thmmain} corresponds to the
optimal wealth process for the expected logarithmic utility
maximization problem under the probability $\qprob$. (This follows by
formally applying first-order conditions for log-optimality and
deriving the ``num\'eraire property'' of log-optimal portfolios---extensive
discussion in the special case of financial models driven by a
finite-dimensional semimartingale integrator can be found in \cite
{MR2335830}.) As can be seen from the proof of Theorem~\ref{thmutilmax}
in Section~\ref{subsecproofofKS}, any optimal process
stemming from utility maximization problems can be regarded as the
log-optimal wealth (more precisely, a multiple of the num\'eraire portfolio
in $\X$) under an auxiliary probability measure that is equivalent to
$\prob$. The idea is certainly not new---for example, in the work of
Kramkov and S\^irbu~\cite{MR2260066,MR2288717}, such changes of num\'eraire
and probability are utilized in questions related to sensitivity
analysis of the expected utility maximization problem as well as
utility indifference prices.
\end{rem}
%
\begin{rem}
Suppose that $\X$ is a wealth-process set such that condition $\mathrm{NA}_1$
holds. In view of Theorem~\ref{thmmain}, condition $\mathrm{NA}_1$ also
holds for the wealth-process set $\oxs$. Indeed, the existence of a
strictly positive $\hX\in\oxs$ such that $\expec[ X_T / \hX_T
] \leq1$ holds for all $X \in\oxs$ can be easily seen to imply
that no arbitrage of the first kind can exist in the market with
wealth-process set~$\oxs$.
\end{rem}

\begin{rem} \label{remfindimintegrator}
Suppose that $\X$ is the wealth-process set generated by nonnegative
stochastic integrals with respect to a finite-dimensional
semimartingale integrator. Then, $\X$ is already closed in the Emery
topology. (The ideas behind the proof of the last claim are present in
M\'emin's work~\cite{MR568256}---see also~\cite{MR2335830},
discussion after
Theorem 4.4, as well as~\cite{Cz-Sch}.) In this special
case, more elaborate versions of Theorem~\ref{thmmain} appear in \cite
{Kar11} and~\cite{Taka10}: condition $\mathrm{NA}_1$ implies that for any
$\qprob\sim\prob$ there exists a strictly positive $\hX^{\qprob}
\in
\X$ such that $X / \hX^\qprob$ is a local $\qprob$-martingale for all
$X \in\X$. Furthermore, the results of~\cite{MR1381678} imply that for
each \textit{maximal} strictly positive wealth process $\hX\in\X$,
there exists $\qprob\sim\prob$ such that $X / \hX$ is a local
$\qprob
$-martingale for all $X \in\X$.
\end{rem}
%
\begin{rem}
Theorem~\ref{thmmain}---which is the basis for proving Theo-\break rem~\ref
{thmutilmax}---underlies the need for assuming that wealth remains
nonnegative; indeed,\vadjust{\goodbreak} the concept of num\'eraire portfolio is only available
for collections of nonnegative processes. The supermartingale property
of properly discounted processes is not suitable to describe optimality
when wealth may become negative. It would be \mbox{interesting} to explore
whether a theory parallel to the one presented here can be developed
for wealth-process sets when processes are not constrained to remain
nonnegative. Naturally, different conditions will be required from a
wealth-process set in such case; for example, an additive analogue of
the multiplicative fork-convexity property of Definition~\ref
{defnwealth-process} may be more appropriate. Such a project will certainly
require different tools than the ones used here and is beyond the scope
of this paper.
\end{rem}

\section{Proofs} \label{secproofs}

\subsection{Some modes of convergence}

Let $\mathbb{L}^0$ be the space of $\F$-measurable $\prob$-a.s.
finitely-valued random variables. For $g
\in\mathbb{L}^0$, define $ \lceil g \rceil_\prob\dfn\expecp[1 \wedge|g|
]$. The metric $(g, g') \mapsto\lceil g - g' \rceil_\prob$ on
$\mathbb{L}^0$ induces the topology of convergence in $\prob$-measure.
We simply write $\mathop{\prob\mbox{-}\lim}_{n \to\infty}g^n = g$
whenever $\lim_{n \to \infty} \lceil g^n - g \rceil_\prob= 0$. We use
$\lzp$ to denote the set of $g \in\mathbb{L}^0$ with $\prob[g \geq 0 ]
= 1$.

For a c\`adl\`ag process $X$, define $X^* \dfn\sup_{t \in[0, \cdot]}
|X_t|$; then, define $ \lceil X \rceil_{\mathsf{u} \prob
} \dfn\lceil X^*_T \rceil_\prob$. The metric $(X, X')
\mapsto\lceil X - X' \rceil_{\mathsf{u} \prob}$
induces the topology of uniform (on $[0, T]$)
convergence in $\prob$-measure on the space of c\`adl\`ag processes. We
write $\mathsf{u}\mathop{\prob\mbox{-}\lim}_{n \to\infty}X^n =
X$ when $\lim_{n \to\infty} \lceil X^n - X \rceil_{\mathsf{u} \prob} =
0$. With the
previous notation, note that $ \lceil X \rceil_{\Sem} =
\sup_{\eta\in\Pre_1}
\lceil\eta\cdot X \rceil_{\mathsf{u} \prob}$ holds
for $X \in\Sem$---in particular, since
considering $\eta\equiv1$ gives $ \lceil X \rceil_{\mathsf{u} \prob}
\leq\lceil X \rceil_{\Sem}$ for $X
\in
\Sem$, $\Sem$-convergence implies $\up$-convergence.

Finally, we introduce yet another mode of convergence. Say that a
sequence of nonnegative c\`adl\`ag processes $(X^n)_{\nin}$ \textit
{Fatou-converges} to a nonnegative c\`adl\`ag process $X$, and write
$\mathsf{F}\mbox{-}\lim_{n \to\infty}X^n = X$, if there exists a
countably dense set $\Time\subseteq
[0, T]$ with $T \in\Time$ such that, $\prob$-a.s.,
\[
X_t = \liminf_{\Time\ni s \downarrow t} \Bigl(\liminf_{n \to\infty
}
X_s^n \Bigr) = \limsup_{\Time\ni s \downarrow t} \Bigl(
\limsup_{n \to\infty} X_s^n \Bigr) \qquad\mbox{for all } t
\in[0, T].
\]
(For $t = T$ the last equality should be read as $X_T = \liminf_{n \to
\infty} X_T^n =\break \limsup_{n \to\infty} X_T^n$.)

\begin{rem}
Fatou-convergence certainly lacks elegance compared to the previous
modes of convergence. However, it proves extremely useful in the theory
of mathematical finance, as was made clear in \cite
{MR1469917,MR1722287} and~\cite{MR1883202}, to name a few. The main
reason for
its usefulness is a ``convex compactness'' property that allows to
obtain existence of optimal wealth processes in the Fatou-closure (the
set of all possible limits in the Fatou sense) of a wealth-process set
for concave maximization problems. Indeed, as stated in Lemma~\ref{lemfcc}
(which follows from~\cite{MR1469917}, Lemma 5.2(1), and a
change-of-num\'eraire argument), if $\X$ is a wealth-process set such that
$\mathrm{NA}_1$ holds, any $\X$-valued sequence\vadjust{\goodbreak} $(X^n)_{\nin}$ has a sequence
of forward convex combinations that is Fatou-convergent. Although
convenient, this ability to easily find Fatou-convergent sequences in
wealth-process sets has the undesirable implication that the
Fatou-closure of a wealth-process set tends to be quite large, making
the corresponding limits difficult to justify from a financial
viewpoint. In fact, Fatou-closures contain ``wealth processes'' that
fail to be maximal, in the sense that they allow for free disposal of
wealth---Section~\ref{subsecpreliminariesforproofofKS} offers a
better understanding of such issues. However, as it turns out,
``optimal'' elements in the Fatou-closure, which are exactly the num\'eraires
mentioned in Theorem~\ref{thmmain}, can be approximated also in the
Emery topology. As already mentioned in Remark
\ref{remfindimintegrator}, when $\X$ is the wealth-process set
generated by nonnegative stochastic integrals with respect to a
finite-dimensional semimartingale integrator, it is established in
\cite{MR1381678} that all strictly positive maximal processes are
actually num\'eraire portfolios under a suitable equivalent change of
probability. However, in the case of possible constraints on
investment, it may happen that maximal elements do not correspond to
num\'eraire portfolios---for an example in a one time-period model, see
\cite{Kar11b}, Section 1.3.
\end{rem}

\subsection{\texorpdfstring{Preliminaries toward proving Theorems
\protect\ref{thmsmarts} and \protect\ref{thmmain}}
{Preliminaries toward proving Theorems 1.3 and 1.7}}

We start with an auxiliary result.
%
\begin{lem}\label{lemNA1sameaspbdd}
Suppose that $\X$ is a wealth-process set. Then, condition
$\mathrm{NA}_1$ holds if and only if $\lim_{\ell\to\infty}
(\sup_{(X, t) \in\X\times[0, T]} \prob[X_t > \ell]
) = 0$, that is, when the
collection $ \{X_t \such X \in\X, t \in[0, T] \}$ of random
variables is bounded in $\prob$-measure.
\end{lem}
\begin{pf}
The proof of the fact that condition $\mathrm{NA}_1$ holds if and only if
$ \{X_T \such X \in\X\}$ is bounded in $\prob$-measure
follows \textit
{mutatis mutandis} from~\cite{MR2832419}, proof of Proposition 1.1. It
only remains to show that boundedness in $\prob$-measure of $ \{
X_T \such X \in\X\}$ implies the stronger boundedness in
$\prob$-measure of
$ \{X_t \such X \in\X, t \in[0, T] \}$. Fix some
strictly positive
$\chi\in\X$, and define $\kappa\in\lzp$ via $\kappa\dfn\sup_{t
\in[0, T]} \chi_t / \chi_T$. For $(X, t) \in\X\times[0, T]$, the
fork-convexity of $\X$ implies that $X_t (\chi_T / \chi_t)$ is equal to
$X'_T$ for some $X' \in\X$. It follows that for any $(X, t) \in\X
\times[0, T]$ there exists $X' \in\X$ such that $X_t \leq\kappa
X'_T$. Since $ \{X_T \such X \in\X\}$ is bounded in
$\prob$-measure
and $\kappa\in\lzp$, it follows that $ \{X_t \such X \in\X,
t \in[0, T] \}$ is bounded in $\prob$-measure as well.
\end{pf}

For a wealth-process set $\X$, let $\overline{\X}{}^{\mathsf{F}}$
denote the set of all possible
limits of Fatou-convergent sequences of $\X$. We state and prove a
result that will help establish both Theorems~\ref{thmsmarts} and
\ref{thmmain}. (Note the similarity between the statements of
Lemma~\ref{lemweakerthanmain} and Theorem~\ref{thmmain}.)
%
\begin{lem}\label{lemweakerthanmain}
Suppose that $\X$ is a wealth-process set and that condition
$\mathrm{NA}_1$ is in force. Then, for all $\qprob\sim\prob$ there
exists a
strictly positive $\hX^\qprob\in\overline{\X}{}^{\mathsf{F}}$ with
\mbox{$\hX^\qprob_0 \geq1$}, such
that $X / \hX^\qprob$ is a $\qprob$-supermartingale for all $X \in
\overline{\X}{}^{\mathsf{F}}$.\vadjust{\goodbreak}
\end{lem}
\begin{pf}
We shall give the proof for the case $\qprob= \prob$ and suppress the
superscript ``$\prob$'' from notation; the proof for the general case
follows in exactly the same way.

Let $\Time$ be a countable dense subset of $[0, T]$ with $ \{0,
T \}
\subseteq\Time$. Recalling Lem\-ma~\ref{lemNA1sameaspbdd}, it
follows exactly as in~\cite{Kar10}, proof of Theorem 2.3, that there
exists an $\X$-valued sequence $(X^n)_{\nin}$ such that:
\begin{longlist}[(a)]
\item[(a)]$\tX_s \dfn\mathop{\prob\mbox{-}\lim}_{n \to\infty}X^n_s$
exists and satisfies $\prob[ \tX_s
> 0 ] = 1$ for all $s \in\Time$; and\vspace*{2pt}
\item[(b)] for all $X \in\X$, $ (X_s / \tX_s )_{s \in\Time}$ is a
$\prob$-supermartingale with respect to the filtration $(\F_s)_{s \in
\Time}$.
\end{longlist}
Using a diagonalization argument and passing to a subsequence if
necessary, we may strengthen $\tX_s = \mathop{\prob\mbox{-}\lim
}_{n \to\infty}X^n_s$ for all $s \in
\Time
$ into that $\prob[ \lim_{n \to\infty}X^n_s = \tX_s$, for all
$s \in
\Time] = 1$. Furthermore, the fact that $X_0 = 1$ for all $X \in
\X
$ coupled with property (b) above gives that $\expecp[ X_s / \tX_s ]
\leq1$ holds for all $X \in\X$ and $s \in\Time$.

Fix a strictly positive semimartingale $X \in\X$. Since the process $ (
X_s/ \tX_s )_{s \in\Time}$ is a nonnegative $\prob$-supermartingale
with respect to the filtration $(\F_s)_{s \in\Time}$, it follows that
$\prob[ \inf_{s \in\Time} \tX_s > 0 ] = 1$. For each $t \in[0, T]$,
define $\hX_t \dfn\lim_{\Time\ni s \downarrow t} \tX_s$; the
$\prob$-a.s. existence of this limit is ensured by the nonnegative
supermartingale convergence theorem. (Note that $\prob[\hX_t < \infty]
= 1$ holds since Lemma~\ref{lemNA1sameaspbdd} implies that the closure
in $\prob$-measure of $ \{X_s \such X \in\X, s \in[0, T] \}$, to which
$\hX_t$ belongs, is bounded in $\prob$-measure.) Since the filtration
$\bF$ satisfies the usual hypotheses, it follows that $\hX$ (viewed as
a process) has an adapted c\`adl\`ag version, which we shall be using
from now on; then, $\mathsf{F}\mbox{-}\lim_{n \to\infty}X^n = \hX$.
Furthermore, $\prob[ \inf_{s \in\Time} \tX_s > 0 ] = 1$ implies that
$\prob[ \inf_{t \in[0, T]} \hX_t > 0] = 1$, that is, that $\hX$ is
strictly positive. The fact that $\expec[ X_s / \tX_s ] \leq1$, for
all $s \in\Time$ and Fatou's lemma give $\expec[ X_t / \hX_t ] \leq1$
for all $t \in[0, T]$. In particular, $1 / \hX_0 = \expec[ X_0 / \hX_0
] \leq1$, that is, $\hX_0 \geq1$.\vspace*{1pt}

It only remains to show that $X/ \hX$ is a $\prob$-supermartingale for
all $X \in\overline{\X}{}^{\mathsf{F}}$. In view of the conditional
version of Fatou's lemma,
it suffices to show that $X/ \hX$ is a $\prob$-supermartingale for all
$X \in\X$. Initially fix $X$ being strictly positive. Let $t \in[0,
T]$, $s \in[0, t]$ and $A \in\F_s$. Consider two $\Time$-valued
sequences $(s_n)_{\nin}$ and $(t_n)_{\nin}$ such that $\downarrow
\lim_{n \to\infty}
s_n = s$, $\downarrow\lim_{n \to\infty}t_n = t$, and $s_n \leq t_n$
for all $\nin
$. Since $A \in\F_{s_n}$ for all $\nin$, property (b) above gives
\[
\expec\biggl[\frac{ \tX_{s_n} X_{t_n} }{ X_{s_n} \tX_{t_n}} \indic_A
\biggr] \leq\prob[A]
\]
for all $\nin$. Taking $n \to\infty$ and using Fatou's lemma, we obtain
\[
\expec\biggl[\frac{ \hX_{s} X_{t} }{ X_{s} \hX_{t}} \indic_A \biggr]
\leq\prob[A].
\]
As $t \in[0, T]$, $s \in[0, t]$ and $A \in\F_s$ are arbitrary, the
last inequality shows that $X/ \hX$ is a $\prob$-supermartingale. The
final step is to remove the assumption that $X$ is strictly positive.
Pick any $X \in\X$ and a strictly positive $X' \in\X$. For all
$\nin
$, define the strictly positive process $X^n \dfn(1 - n^{-1}) X +
n^{-1} X'$, which is a wealth process in $\X$. It follows that $X^n /
\hX$ is a nonnegative $\prob$-supermartingale for all $\nin$. Using the
conditional version of Fatou's lemma, it follows that $X / \hX$ is a
nonnegative $\prob$-supermartingale, which concludes the argument.
\end{pf}

\subsection{\texorpdfstring{Proof of Theorem \protect\ref{thmsmarts}}
{Proof of Theorem 1.3}}

Fix a strictly positive semimartingale $X' \in\X$ and (in view of
Lemma~\ref{lemweakerthanmain}) a strictly positive $\hX\in\overline
{\X}{}^{\mathsf{F}}$
such that $X / \hX$ is a $\prob$-supermartingale for all $X \in\X$.
Pick any $X \in\X$ and write $X = (X / \hX) (\hX/ X') X'$. The
process $X / \hX$ is a c\`adl\`ag supermartingale, therefore a
semimartingale. As $X' \in\Sem$, $X \in\Sem$ will follow as soon as
$(\hX/ X') \in\Sem$ is established. The last follows upon noticing
that $\hX/ X' = 1/ (X' / \hX)$ and using It\^o's formula with the
function $(0, \infty) \ni x \mapsto1/x \in(0, \infty)$ on the
strictly positive semimartingale $X' / \hX$.

\subsection{Convergence in the Emery topology} \label{subsecsemconv}

Below, we collect the essential results regarding convergence in the
Emery topology that shall be needed for the proof of Theorem \ref
{thmmain}. We provide full details for the convenience of the reader;
however, versions of some of them have appeared previously---for
example, see the original paper~\cite{MR544800}.
%
\begin{conv} \label{convstopinf}
In several occasions until the end of Section
\ref{subsecproofofthmmain}, we define stopping times as first passage times
of processes
in certain sets. On the event that the process never enters the
specific set up to time $T$, the stopping time is defined by convention
equal to $\infty$.
\end{conv}

The first result contains a convenient necessary and sufficient
condition for $\Sem$-convergence.
%
\begin{lem} \label{lemnecandsuffforsem-conv}
Let $(X^n)_{\nin}$ be a sequence in $\Sem$. Then, $\mathcal{S}\mbox
{-}\lim_{n \to\infty}X^n = 0$ holds
if and only if for all $\Pre_1$-valued sequences $(\eta^n)_{\nin}$,
$\mathop{\prob\mbox{-}\lim}_{n \to\infty} (\eta^n \cdot
X^n )_T = 0$ holds.
\end{lem}
\begin{pf}
By definition, $\mathcal{S}\mbox{-}\lim_{n \to\infty}X^n = 0$
implies $\mathop{\prob\mbox{-}\lim}_{n \to\infty} (\eta^n
\cdot X^n )_T
= 0$ whenever $(\eta^n)_{\nin}$ is a $\Pre_1$-valued sequence. Now,
assume the latter condition and, by way of contradiction, that
$\mathcal{S}\mbox{-}\lim_{n \to\infty}X^n
= 0$ fails. Passing to a subsequence if necessary, one can find
$\varepsilon> 0$ and a $\Pre_1$-valued sequence $(\theta^n)_{\nin}$ such
that $\prob[(\theta^n \cdot X^n)^*_T > \varepsilon] >
\varepsilon$ for
all $\nin$. For each $\nin$, define the stopping time $\tau^n \dfn
\inf
\{t \in[0, T] \such\llvert\theta^n \cdot X^n\rrvert_t > \varepsilon\}$. With
$\eta^n \dfn\theta^n \indic_{\dbra{0, \tau^n \wedge T}}$, $(\eta
^n)_{\nin}$ is $\Pre_1$-valued sequence, and $\mathop{\prob\mbox
{-}\lim}_{n \to\infty} (\eta^n \cdot X^n )_T = 0$
fails. We reached a contradiction, which means that $\mathcal{S}\mbox
{-}\lim_{n \to\infty}
X^n = 0$ holds.
\end{pf}

We introduce some notation that will be used in all that follows. For
$X \in\Sem$, $X_-$ denotes its left-continuous version, with the
understanding that $X_{0-} = 0$. We define $\Delta X \dfn X - X_-$. The
quadratic covariation process between $X \in\Sem$ and $Y \in\Sem$ is
$[X, Y] \dfn X Y - X_- \cdot Y - Y_- \cdot X$. (Note that $[X, Y]_0 =
X_0 Y_0$.) Furthermore, $\var(X)$ denotes the first-variation process
of $X \in\Sem$.
%
\begin{rem} \label{remdblsubsec}
During the remainder of Section~\ref{subsecsemconv}, some proofs
make use of the following \textit{double subsequence} trick. Suppose that
\textit{any} subsequence of a given a sequence of random variables has a
further subsequence that converges in $\prob$-measure to zero. As
convergence in $\prob$-measure comes from a metric topology, it follows
that the whole sequence has to converge to zero in $\prob$-measure.
\end{rem}

The next result discusses sufficient conditions for $\Sem$-convergence
that will be used in the main text.
%
\begin{prop} \label{propsuffforsem-conv}
If $(X^n)_{\nin}$ is a sequence of semimartingales,
\[
\mathcal{S}\mbox
{-}\lim_{n \to\infty}X^n = 0
\]
holds in all of the following three cases:
\begin{itemize}
\item$\lim_{n \to\infty}\prob[(X^n)_T^* > 0] = 0$.
\item Each $X^n$ is a process of finite variation, and $\mathop{\prob
\mbox{-}\lim}_{n \to\infty}\var
(X^n)_T = 0$.
\item Each $X^n$ is a local martingale with $|\Delta X^n| \leq C$,
where $C \in\Real_+$ does not depend on $\nin$, and $\mathop{\prob
\mbox{-}\lim}_{n \to\infty}[X^n,
X^n]_T = 0$.
\end{itemize}
\end{prop}
\begin{pf}
We treat each case separately below.

First, assume that $\lim_{n \to\infty}\prob[(X^n)_T^* > 0] = 0$. On
the event
$ \{(X^n)_T^* = 0 \}$ we have $\eta^n \cdot X^n = 0$ for
all $\eta^n \in
\Pre_1$ in view of~\cite{MR2273672}, Chapter IV, Theorem 26. Then the
result follows from Lemma~\ref{lemnecandsuffforsem-conv}.

Now, assume that each $X^n$ is a process of finite variation, and
\[
\prob\mbox{-}\lim_{n \to\infty}\var\bigl(X^n\bigr)_T = 0
\]
holds. For $\eta^n \in\Pre_1$ we have $\llvert(\eta^n \cdot
X^n)_T\rrvert\leq\var (X^n)_T$---then, Lemma \ref
{lemnecandsuffforsem-conv} allows us to conclude.

Finally, assume that each $X^n$ is a local martingale with $|\Delta
X^n| \leq C$ for $C \in\Real_+$, and that $\mathop{\prob\mbox
{-}\lim}_{n \to\infty}[X^n, X^n]_T = 0$.
Let $(\eta^n)_{\nin}$ be a $\Pre_1$-valued sequence and set $M^n =
\eta^n \cdot X^n$ for $\nin$. We need to show that $\mathop{\prob\mbox
{-}\lim}_{n \to\infty}M^n_T = 0$. Note
that $|\Delta M^n| = |\eta^n \Delta X^n| \leq C$ and $[M^n, M^n] =
|\eta^n|^2 \cdot[X^n, X^n] \leq[X^n, X^n]$ so that $\mathop{\prob\mbox
{-}\lim}_{n \to\infty}[M^n, M^n]_T =
0$. Let $(M^{n_k})_{\kin}$ be a subsequence of $(M^{n})_{\nin}$ such
that $\prob[[M^{n_k}, M^{n_k}]_T > 1 / 2^{k} ] \leq1 /
2^k$ holds
for all $k \in\Natural$; then, by the first Borel--Cantelli lemma it
follows that $A \dfn\sum_{\kin} [M^{n_k}, M^{n_k}]$ is a finite
nondecreasing adapted process. For $m \in\Natural$, define $\tau_m
\dfn\inf\{t \in[0, T] \such A_t \geq m \}$. Then, $[M^{n_k},
M^{n_k}]_{\tau_m} \leq A_{\tau_m -} + (\Delta M^{n_k})_{\tau_m}^2
\leq
m + C^2$ holds for all $\kin$ and $m \in\Natural$. Therefore, using
the well-known $\mathbb{L}^2$-isometry for square-integrable
martingales and the dominated convergence theorem, we obtain
\[
\lim_{k \to\infty}\expec\bigl[\bigl|M^{n_k}_{\tau_m \wedge T}\bigr|^2
\bigr] = \lim_{k \to\infty}\expec\bigl[\bigl[M^{n_k}, M^{n_k}
\bigr]_{\tau_m
\wedge T} \bigr] = 0.
\]
This implies that
\[
\prob\mbox{-}\lim_{k \to\infty
}M^{n_k}_{\tau_m \wedge T} = 0
\]
and, in turn,\vspace*{1pt} that $\mathop{\prob\mbox{-}\lim}_{k \to\infty} (M^{n_k}_T
\indic_{ \{\tau_m = \infty\}} )= 0$. The fact that $\prob[\bigcup_{m
\in\Natural} \{\tau_m = \infty\} ] = 1$ implies that
$\mathop{\prob\mbox{-}\lim}_{k \to\infty }M^{n_k}_T= 0$. Up to now we
have shown that there exists a subsequence of $ (M^{n}_T )_{\nin}$ that
converges in $\prob$-measure to zero. The same argument shows that
\textit{any} subsequence of $ (M^{n}_T )_{\nin}$ has a further
subsequence that converges in $\prob$-measure to zero. By the double
subsequence trick mentioned in Remark~\ref{remdblsubsec}, it follows
that $\mathop {\prob\mbox{-}\lim}_{n \to\infty} M^n_T = 0$, which
concludes the argument.
\end{pf}
%
\begin{rem}
Let $(X^n)_{\nin}$ be a sequence of local martingales such that
$\mathop{\prob\mbox{-}\lim}_{n \to\infty}
[X^n, X^n]_T = 0$ holds. In the case where there does \textit{not} exist
any $C \in\Real_+$ with $|\Delta X^n| \leq C$ holding for all $\nin$,
$\mathcal{S}\mbox{-}\lim_{n \to\infty}X^n = 0$ may fail. For
example, consider a probability space
$(\Omega, \F, \prob)$ that affords a collection $ \{\tau_n
\such\nin\}$ of independent (under $\prob$) random
variables such that $\prob
[\tau_n > t ] = \exp(-t/n)$ for $t \in\Real_+$. Define
$(\F_t)_{t
\in
[0, T]}$ as (the restriction on $[0, T]$ of) the usual augmentation of
the smallest filtration that makes all random times in the collection
$ \{\tau_n \such\nin\}$ stopping times. Then, for each
$\nin$, define
a martingale $X^n$ via the formula $X^n_t = n \indic_{ \{\tau^n
\leq t \}} - \tau_n \wedge t$ for $t \in[0, T]$. [It is
straightforward to
check that each~$X^n$, $\nin$, is a martingale in its own filtration;
then, the independence of the random variables in $ \{\tau_n
\such\nin\}$ implies that $X^n$ is also a martingale in the
larger filtration
$(\F_t)_{t \in[0, T]}$, for all $\nin$.] In this case, $[X^n, X^n]_T =
n^2 \indic_{ \{\tau^n \leq T \}}$ for all $\nin$; as
$\lim_{n \to\infty}\prob
[\tau^n \leq T ] = 0$, $\mathop{\prob\mbox{-}\lim}_{n
\to\infty}[X^n, X^n]_T = 0$ holds. However,
$\mathop{\prob\mbox{-}\lim}_{n \to\infty}
X^n_T = - T$, which of course implies that $\mathcal{S}\mbox{-}\lim_{n
\to\infty}X^n = 0$ fails.
\end{rem}

The two last results of Section~\ref{subsecsemconv} concern
stability of $\Sem$-convergence.
%
\begin{lem} \label{lemstochintegralconv}
Let $\mathcal{S}\mbox{-}\lim_{n \to\infty}X^n = X$ and $(Y^n)$ be
a sequence of adapted c\`adl\`ag
processes such that $\mathsf{u}\mathop{\prob\mbox{-}\lim}_{n \to
\infty}Y^n = Y$. Then, $\mathcal{S}\mbox{-}\lim_{n \to\infty
}(Y^n_- \cdot X^n) = Y_-
\cdot X$.\vadjust{\goodbreak}
\end{lem}
\begin{pf}
Upon writing $Y^n_- \cdot X^n - Y_- \cdot X = Y_- \cdot(X^n - X) +
(Y^n - Y)_- \cdot X + (Y^n - Y)_- \cdot(X^n - X)$, it suffices to
treat three special cases: (i) when $Y^n = Y$ for all $\nin$ and
$\mathcal{S}\mbox{-}\lim_{n \to\infty}
X^n = 0$, (ii) when $\mathsf{u}\mathop{\prob\mbox{-}\lim}_{n \to
\infty}Y^n = 0$ and $X^n = X$ for all $\nin$ and
(iii) when $\mathsf{u}\mathop{\prob\mbox{-}\lim}_{n \to\infty
}Y^n = 0$ and $\mathcal{S}\mbox{-}\lim_{n \to\infty}X^n = 0$ both hold.

First, assume case (i): $Y^n = Y$ for all $\nin$ and $\mathcal
{S}\mbox{-}\lim_{n \to\infty}X^n = 0$. For $\kin$, define $\tau_k
\dfn\inf\{t \in[0, T] \such|Y_t| > k \}$. Let $(\eta^n)_{\nin}$ be a
$\Pre_1$-valued sequence and set $\theta^{k, n} \dfn\eta^n (Y_- / k)
\indic_{\dbra{0, \tau_k \wedge T}}$. Noting\vspace*{1pt} that
$(\theta^{k, n})_{\nin}$ is a $\Pre_1$-valued sequence, it follows that
$\mathop{\prob\mbox{-}\lim}_{n \to\infty} (\eta^n \cdot(Y_- \cdot X^n)
)_{\tau_k \wedge T} =\break k \mathop{\prob\mbox{-}\lim}_{n \to\infty}
(\theta^{k, n} \cdot X^n )_T = 0$. Therefore,
$\mathop{\prob\mbox{-}\lim}_{n \to\infty} (\eta^n \cdot(Y_- \cdot X^n)
)_T \indic_{ \{\tau_k = \infty\}} = 0$. Since it holds that
$\prob[\bigcup_{\kin} \{\tau_k = \infty\} ] = 1$, we obtain
$\mathop{\prob\mbox{-}\lim}_{n \to\infty} (\eta^n \cdot(Y_- \cdot X^n)
)_T = 0$. As the $\Pre_1$-valued sequence $(\eta^n)_{\nin}$ was
arbitrary, Lem\-ma~\ref{lemnecandsuffforsem-conv} implies that
$\mathcal{S}\mbox{-}\lim_{n \to\infty}(Y_- \cdot X^n) = 0$.

Now, assume case (ii): $\mathsf{u}\mathop{\prob\mbox{-}\lim}_{n
\to\infty}Y^n = 0$ and $X^n = X$ for all $\nin$.
For an arbitrary $\Pre_1$-valued sequence $(\eta^n)_{\nin}$, we shall
show that $\mathop{\prob\mbox{-}\lim}_{n \to\infty} (\eta^n
\cdot(Y^n_- \cdot X) )_T = 0$. Pick a
subsequence $(Y^{n_k})_{\kin}$ such that $\xi\dfn\sum_{\kin}
|Y^{n_k}|$ is a real-valued c\`adl\`ag process. The facts that $\mathop
{\prob\mbox{-}\lim}_{k \to\infty}
(\eta^{n_k} Y^{n_k}_-)^* = 0$, $\xi_-$ is $X$-integrable (since $\xi_-$
is locally bounded) and $|\eta^{n_k} Y^{n_k}_-| \leq\xi_-$ for all
$\kin$, coupled with the dominated convergence theorem for stochastic
integrals, imply that $\mathop{\prob\mbox{-}\lim}_{k \to\infty
} ((\eta^{n_k} Y^{n_k}_-) \cdot X )_T =
0$, that is, $\mathop{\prob\mbox{-}\lim}_{k \to\infty} (\eta^{n_k} \cdot
(Y^{n_k}_- \cdot X ) )_T =
0$. Up to now we have shown that there exists a subsequence of $
( (\eta^n \cdot(Y^n_- \cdot X) )_T )_{\nin}$ that
converges in
$\prob
$-measure to zero. The same argument shows that \textit{any} subsequence
of $ ( (\eta^n \cdot(Y^n_- \cdot X) )_T )_{\nin
}$ has a further
subsequence that converges in $\prob$-measure to zero. The double
subsequence trick of Remark~\ref{remdblsubsec} allows us to conclude
that $\mathop{\prob\mbox{-}\lim}_{n \to\infty} (\eta^n
\cdot(Y^n_- \cdot X) )_T = 0$. As the
sequence $(\eta^n)_{\nin}$ was arbitrary, Lem\-ma~\ref
{lemnecandsuffforsem-conv} implies that $\mathcal{S}\mbox{-}\lim_{n
\to\infty}Y^n \cdot X = 0$.

Finally, assume case (iii): $\mathsf{u}\mathop{\prob\mbox{-}\lim
}_{n \to\infty}Y^n = 0$ and $\mathcal{S}\mbox{-}\lim_{n \to\infty
}X^n = 0$ for all
$\nin$. In view of Lemma~\ref{lemnecandsuffforsem-conv}, we only
need to show that $\mathop{\prob\mbox{-}\lim}_{n \to\infty}
(\eta^n \cdot(Y^n_- \cdot X^n) )_T = 0$
for an arbitrary $\Pre_1$-valued sequence $(\eta^n)_{\nin}$. Similarly
to case~(ii), pick a subsequence $(Y^{n_k})_{\kin}$ such that $\xi
\dfn
\sum_{\kin} |Y^{n_k}|$ is a real-valued c\`adl\`ag process. For $m
\in
\Natural$, define $\tau_m \dfn\inf\{t \in[0, T] \such|\xi_t| > m \}$.
For $m \in\Natural$ and $\kin$, set $\theta^{m,
k} \dfn\eta^{n_k} (Y^{n_k}_- / m) \indic_{\dbra{0, \tau_m \wedge T}}$. As
$(\theta^{m, k})_{\kin}$ is $\Pre_1$-valued, we have $\mathop{\prob\mbox
{-}\lim}_{k \to\infty} (\eta^{n_k} \cdot(Y^{n_k}_- \cdot
X^{n_k}) )_{\tau_m \wedge T} = m \mathop{\prob\mbox{-}\lim
}_{k \to\infty} (\theta^{m, k} \cdot X^{n_k} )_T = 0$.
Therefore, for all $m \in
\Natural
$, $\mathop{\prob\mbox{-}\lim}_{k \to\infty} (\eta^{n_k}
\cdot(Y^{n_k}_- \cdot X^{n_k}) )_T \indic_{ \{\tau_m = \infty\}} = 0$
holds. Since $\prob
[\bigcup_{m \in\Natural} \{\tau_m = \infty\} ]
= 1$, we obtain that $\mathop{\prob\mbox{-}\lim}_{k \to\infty
} (\eta^{n_k} \cdot(Y_- \cdot X^{n_k}) )_T = 0$. We have
shown that there
exists a subsequence of $ ( (\eta^n \cdot(Y^n_- \cdot
X^n) )_T )_{\nin}$ that converges in $\prob$-measure to
zero. The same
argument shows that \textit{any} subsequence of $ ( (\eta^n
\cdot(Y^n_- \cdot X^n) )_T )_{\nin}$ has a further
subsequence that converges
in $\prob$-measure to zero. By the double subsequence trick of Remark
\ref{remdblsubsec}, it follows that $\mathop{\prob\mbox{-}\lim}_{n
\to\infty} (\eta^n \cdot(Y^n_- \cdot X^n) )_T = 0$.
Then, another invocation of Lemma \ref
{lemnecandsuffforsem-conv} implies that $\mathcal{S}\mbox{-}\lim_{n
\to\infty} (Y^n_- \cdot X^n ) = 0$.
\end{pf}
%
\begin{prop} \label{propstabilityofsem-conv}
$\!\!\!$Let $\mathcal{S}\mbox{-}\lim_{n \to\infty}X^n = X$ and $\mathcal
{S}\mbox{-}\lim_{n \to\infty}Y^n = Y$. Then, it further holds that
$\mathcal{S}\mbox{-}\lim_{n \to\infty}
[X^n, Y^n] = [X, Y]$ and \mbox{$\mathcal{S}\mbox{-}\lim_{n \to\infty
}(X^n Y^n) = X Y$}.
\end{prop}
\begin{pf}
We shall establish below that $\mathcal{S}\mbox{-}\lim_{n \to\infty
}[X^n, Y^n] = [X, Y]$; then, $\mathcal{S}\mbox{-}\lim_{n \to\infty}
(X^n Y^n) = X Y$ follows from Lemma~\ref{lemstochintegralconv} and
a use of the integration-by-parts formula.

Using the identity $4 [X^n, Y^n] = [X^n+ Y^n, X^n+ Y^n] - [X^n - Y^n,
X^n - Y^n]$, it follows that it suffices to show that $\mathcal
{S}\mbox{-}\lim_{n \to\infty}X^n = X$
implies $\mathcal{S}\mbox{-}\lim_{n \to\infty}[X^n, X^n] = [X,
X]$. Furthermore, since quadratic
variation processes of semimartingales are of finite variation, the estimate
\begin{eqnarray*}
\var\bigl(\bigl[X^n, X^n\bigr] - [X, X]
\bigr)_T &=& \var\bigl(\bigl[X^n - X, X^n - X
\bigr] + 2 \bigl[X, X^n - X\bigr] \bigr)_T
\\
&\leq&\bigl[X^n - X, X^n - X\bigr]_T\\
&&{} + 2
\sqrt{[X,X]_T} \sqrt{\bigl[X^n - X, X^n - X
\bigr]_T}
\end{eqnarray*}
implies that we only have to establish that, whenever $\mathcal
{S}\mbox{-}\lim_{n \to\infty}X^n = 0$,
$\mathcal{S}\mbox{-}\lim_{n \to\infty}[X^n, X^n] = 0$ holds. In
view of Proposition \ref
{propsuffforsem-conv}, $\mathcal{S}\mbox{-}\lim_{n
\to\infty}[X^n,\break
X^n] = 0$ is equivalent to $\mathop{\prob\mbox{-}\lim}_{n \to
\infty}
[X^n, X^n]_T =
0$. Using $[X^n, X^n] = |X^n|^2 - 2 X^n_- \cdot X^n$ as well as that
$\mathsf{u}\mathop{\prob\mbox{-}\lim}_{n \to\infty}X^n = 0$ and
$\mathsf{u}\mathop{\prob\mbox{-}\lim}_{n \to\infty}(X^n_- \cdot
X^n) = 0$, the latter holding
in view of Lemma~\ref{lemstochintegralconv}, we obtain the result.
\end{pf}

\subsection{\texorpdfstring{Proof of Theorem \protect\ref{thmmain}}{Proof of Theorem 1.7}} \label
{subsecproofofthmmain}

In the course of the proof of Theorem~\ref{thmmain}, we shall
actually assume that $\qprob= \prob$ and use ``$\prob$'' in what
follows for notational simplicity. Of course, this does not entail any
loss of generality whatsoever. (Note that the Emery topology depends on
the probability measure only through its equivalence class.)

Suppose that $\X$ is a wealth-process set and that condition $\mathrm{NA}_1$
is valid. Keeping the notation of Lemma~\ref{lemweakerthanmain},
consider the strictly positive $\hX\equiv\hX^\prob\in\overline{\X
}{}^{\mathsf{F}}$ with
$\hX_0 \geq1$ and such that $X / \hX$ is a $\prob$-supermartingale for all
$X \in\overline{\X}{}^{\mathsf{F}}\supseteq\X$. Pick an $\X
$-valued sequence $(X^n)_{\nin}$
such that $\mathsf{F}\mbox{-}\lim_{n \to\infty}X^n = \hX$; in
particular, $\mathop{\prob\mbox{-}\lim}_{n \to\infty}X^n_T = \hX_T$.
Define $Z^n \dfn X^n / \hX$, which is a nonnegative $\prob
$-supermartingale with $Z^n_0 \leq1$ for all $\nin$. The convergence
$\mathop{\prob\mbox{-}\lim}_{n \to\infty}X^n_T = \hX_T$
translates to $\mathop{\prob\mbox{-}\lim}_{n \to\infty}Z^n_T =
1$. If one can
show that $\mathcal{S}\mbox{-}\lim_{n \to\infty}Z^n = 1$, an
application of Proposition \ref
{propstabilityofsem-conv} shows that $\mathcal{S}\mbox{-}\lim_{n \to
\infty}X^n = \hX$, which will complete
the argument. Therefore, we shall prove below that if a sequence
$(Z^n)_{\nin}$ of nonnegative $\prob$-supermartingales with $Z^n_0
\leq
1$ for all $\nin$ satisfies $\mathop{\prob\mbox{-}\lim}_{n \to
\infty}Z^n_T = 1$, then $\mathcal{S}\mbox{-}\lim_{n \to\infty}Z^n
= 1$. We
prepare the ground with
the following result, which establishes $\up$-convergence. 
In the course of the proofs below, Convention~\ref{convstopinf} will
be used.
%
\begin{lem} \label{lemucpconv}
Suppose that $(Z^n)_{n \in\Natural}$ is a sequence of nonnegative
$\prob$-supermartingales such that $Z^n_0 \leq1$ for all $n \in
\Natural$, as well as $\mathop{\prob\mbox{-}\lim}_{n \to\infty
}Z^n_T = 1$. Then, in fact, $\mathsf{u}\mathop{\prob\mbox{-}\lim
}_{n \to\infty}Z^n = 1$.\vadjust{\goodbreak}
\end{lem}
\begin{pf}
Since $\expec[Z^n_T] \leq1$ for all $n \in\Natural$, $\lim_{n \to
\infty}\expec
[Z_T^n] = 1$ holds by Fatou's lemma. Then,~\cite{MR2722836},
Theorem 5.5.2, implies the uniform integrability of $(Z^n_T)_{n \in
\Natural}$; therefore, $\lim_{n \to\infty}\expec[\llvert
Z_T^n - 1\rrvert ] = 0$.

We shall now show that $\mathop{\prob\mbox{-}\lim}_{n \to\infty
}\sup_{t \in[0, T]} Z_t^n = 1$. Fix
$\varepsilon\in(0, \infty)$ and define the stopping time $\tau^n:=
\inf
\{t \in[0, T] \such Z^n_t > 1 + \varepsilon\}$ for all $n \in\Natural$.
Showing that $\lim_{n \to\infty}\oprob[\tau^n = \infty] = 1$ will
imply that
$\mathop{\prob\mbox{-}\lim}_{n \to\infty}\sup_{t \in[0, T]}
Z^n_t = 1$, since $\varepsilon\in(0, \infty
)$ is arbitrary. Suppose on the contrary (passing to a subsequence if
necessary) that $\lim_{n \to\infty}\oprob[\tau^n = \infty] = 1 -
p$, where $p > 0$.
Then, since $\llvert \expec[Z^n_{T} \indic_{\{ \tau^n =
\infty\}} ] - \oprob[\tau^n = \infty] \rrvert = \llvert
\expec[(Z_T^n - 1) \indic_{\{ \tau^n = \infty\}} ]
\rrvert \leq\expec[|Z_T^n - 1|]$,
and the last quantity converges to zero as $n \to\infty$, we obtain
$\lim_{n \to\infty}\expec[Z^n_{T} \indic_{\{ \tau^n =
\infty\}} ] = 1 - p$.
In turn, this implies
\begin{eqnarray*}
1 &\geq&\limsup_{n \to\infty} \expec\bigl[Z^n_{0}\bigr]
\geq\limsup_{n
\to\infty} \expec\bigl[Z^n_{\tau^n \wedge T}\bigr] \\
&\geq&
\liminf_{n \to
\infty
} \expec\bigl[ Z^n_{\tau^n}
\indic_{\{ \tau^n \leq T \}} \bigr] + \lim_{n \to\infty} \expec\bigl[
Z^n_{T} \indic_{\{ \tau^n = \infty\}} \bigr]
\\
&\geq& (1 + \varepsilon) p + (1 - p) = 1 + \varepsilon p,
\end{eqnarray*}
which contradicts the fact that $p > 0$. Thus, $\mathop{\prob\mbox
{-}\lim}_{n \to\infty}\sup_{t \in[0,
T]} Z_t^n = 1$ has been shown.

We shall now establish that $\mathop{\prob\mbox{-}\lim}_{n \to
\infty}\inf_{t \in[0, T]} Z_t^n = 1$. Fix
$\varepsilon\in(0, \infty)$, and for each $n \in\Natural$ redefine
$\tau^n:= \inf\{t \in[0, T] \such Z^n_t < 1 - \varepsilon\}$---we only need
to show that $\lim_{n \to\infty}\oprob[\tau^n = \infty] = 1$. The
nonnegative
supermartingale property of $Z^n$ gives that, on $\{ \tau^n \leq T \}$,
where in particular $Z_{\tau^n} \leq1 - \varepsilon$ holds, we have
$\oprob[Z^n_T > 1 - \varepsilon^2 \such\F_{\tau^n}] \leq(1 -
\varepsilon) /
(1 - \varepsilon^2) = 1 / (1 + \varepsilon)$ for all $\nin$. Then,
\[
\oprob\bigl[Z^n_T > 1 - \varepsilon^2\bigr] =
\expec\bigl[ \oprob\bigl[Z^n_T > 1 -
\varepsilon^2 \such\F_{\tau^n}\bigr] \bigr] \leq\oprob\bigl[
\tau^n = \infty\bigr] + \oprob\bigl[\tau^n \leq T\bigr]
\frac{1}{1+\varepsilon}.
\]
Using $\oprob[\tau^n = \infty] = 1- \oprob[\tau^n \leq T]$, rearranging
and taking the inferior limit as $n \to\infty$, we obtain $\liminf_{n
\to\infty} \oprob[\tau^n = \infty] \geq(1 + \varepsilon^{-1} )
\liminf_{n \to\infty} \oprob[Z_T^n > 1 - \varepsilon^2] - \varepsilon^{-1} =
1$, which
shows that $\mathop{\prob\mbox{-}\lim}_{n \to\infty}\inf_{t \in
[0, T]} Z_t^n = 1$. Together with
$\mathop{\prob\mbox{-}\lim}_{n \to\infty}\sup_{t \in[0, T]}
Z_t^n = 1$ that was proved above, the proof
of Lemma~\ref{lemucpconv} is complete.
\end{pf}

Theorem~\ref{thmmain} immediately follows from Propositions \ref
{propsuffforsem-conv},~\ref{propstabilityofsem-conv}, and
the following result.
%
\begin{lem} \label{lemkeyforsem-conv}
Under the assumptions of Lemma~\ref{lemucpconv}, one can write $Z^n
= 1 + A^n - B^n + L^n$ for each $\nin$, where:
\begin{itemize}
\item Each $A^n$ is a semimartingale, and $\lim_{n \to\infty}\prob
[(A^n)_T^* > 0]
= 0$.
\item Each $B^n$ is a predictable, nonnegative and nondecreasing
process, and $\mathop{\prob\mbox{-}\lim}_{n \to\infty}B^n_T = 0$.
\item Each $L^n$ is a local martingale with $|\Delta L^n| \leq4$ and
$\mathop{\prob\mbox{-}\lim}_{n \to\infty}[L^n, L^n]_T = 0$.
\end{itemize}
\end{lem}
\begin{pf}
For $\nin$, define the stopping time $\tau^n:= \inf\{ t \in[0, T]
\such Z_t^n > 2 \}$. Furthermore, for $\nin$ define processes $\zeta^n$
and $A^n$ via $\zeta^n_t = Z^n_{t \wedge\tau^n} - \Delta Z^n_{\tau^n}
\indic_{\{ \tau^n \leq t \}}$ and $A^n_t = (Z^n_{t} - Z^n_{\tau^n -})
\indic_{\{ \tau^n \leq t \}}$ for $t \in[0, T]$. In other words,
$\zeta^n$ is the process $Z^n$ stopped \textit{just before} time $\tau^n$,
while $A^n$ is defined so that \mbox{$Z^n = A^n + \zeta^n$}. Since $\Delta
Z^n_{\tau^n} \geq0$, $\zeta^n$ is a supermartingale and $0 \leq
\zeta^n \leq2$ holds for all $n \in\Natural$. Now, $\lim_{n \to\infty
}\prob[ \tau^n =
\infty] = 1$ holds in view of Lemma~\ref{lemucpconv}; therefore,
$\lim_{n \to\infty}\prob[(A^n)_T^* > 0] = 0$, as required. Since
$\mathsf{u}\mathop{\prob\mbox{-}\lim}_{n \to\infty}Z^n = 1$
and $\mathsf{u}\mathop{\prob\mbox{-}\lim}_{n \to\infty}A^n = 0$,
we obtain $\mathsf{u}\mathop{\prob\mbox{-}\lim}_{n \to\infty
}\zeta^n = 1$. For each $n \in
\Natural$, write $\zeta^n = - B^n + M^n$ for the Doob--Meyer
decomposition of $\zeta^n$, where $B^n$ is predictable, nonnegative and
nondecreasing process and such that $B^n_0 = 0$, and $M^n$ is a
nonnegative local martingale with $M^n_0 = \zeta^n_0 = Z^n_0 \leq1$.
Since $M^n \geq\zeta^n$ and $\mathop{\prob\mbox{-}\lim}_{n \to
\infty}\zeta^n_T = 1$, it necessarily
holds that $\mathop{\prob\mbox{-}\lim}_{n \to\infty}M^n_T = 1$;
otherwise $\limsup_{n \to\infty}
\expec
[M^n_T] > 1$, which is impossible in view of the fact that $M_0^n \leq
1$ and $M^n$ is a nonnegative local $\prob$-martingale for all $\nin$.
Using $\mathop{\prob\mbox{-}\lim}_{n \to\infty}\zeta^n_T = 1$
and $\mathop{\prob\mbox{-}\lim}_{n \to\infty}M^n_T = 1$, we
obtain $\mathop{\prob\mbox{-}\lim}_{n \to\infty}
B^n_T = 0$, which completes the requirements for the sequence
$(B^n)_{\nin}$.

Continuing, a use of Lemma~\ref{lemucpconv} with $(M^n)_{\nin}$ in
place of $(Z^n)_{\nin}$, gives $\mathsf{u}\mathop{\prob\mbox
{-}\lim}_{n \to\infty}M^n = 1$. We define $L^n$ in the
obvious way: $L^n = M^n - 1$; it remains to show that the requirements
for the sequence $(L^n)_{\nin}$ are fulfilled. First, note that $0
\leq\zeta^n \leq2$ implies that $|\Delta\zeta^n| \leq2$; therefore,
$0 \leq\Delta B^n \leq2$, since $\Delta B^n_\tau= - \expec[\Delta
\zeta^n_\tau\such\F_{\tau}] + [\Delta M^n_\tau\such\F_{\tau}]
= -
\expec[\Delta\zeta^n_\tau\such\F_{\tau}]$ holds for all predictable
times $\tau$. This implies that $|\Delta L^n| = |\Delta M^n| \leq
|\Delta\zeta^n| + \Delta B^n \leq4$. It only remains to show that
$\mathop{\prob\mbox{-}\lim}_{n \to\infty}[L^n, L^n]_T = 0$. Fix
$\varepsilon\in(0,\infty)$ and redefine,
for each $\nin$, the stopping time $\tau^n:= \inf\{ t \in[0, T]
\such M_t^n > 1 / \varepsilon\}$. Since $M^n_0 \leq1$ and each $M^n$ is
a nonnegative local $\prob$-martingale, we obtain that $\prob[\tau^n =
\infty] \geq1 - \varepsilon$. Also, note that $\sup_{t \in[0, T]}
|L_{\tau^n \wedge t}| \leq1 + \sup_{t \in[0, T]} M_{\tau^n \wedge t}
\leq5 + 1 / \varepsilon$ for all $\nin$. Coupled with the fact that
$\mathop{\prob\mbox{-}\lim}_{n \to\infty}M^n_{\tau^n \wedge T} =
1$ (recall that $\mathsf{u}\mathop{\prob\mbox{-}\lim}_{n \to
\infty}M^n = 1$) and
the $L^2$-isometry for square-integrable martingales, we obtain
\[
\lim_{n \to\infty}\expec\bigl[\bigl[L^n, L^n
\bigr]_{\tau^n \wedge T} \bigr] = \lim_{n \to\infty}\expec\bigl
[\bigl|L^n_{\tau^n \wedge T}\bigr|^2
\bigr] = \lim_{n \to\infty}\expec\bigl[\bigl|M^n_{\tau^n \wedge T} -
1\bigr|^2 \bigr] = 0.
\]
It follows that $\limsup_{n \to\infty} \prob[[L^n, L^n]_{T} >
\varepsilon] \leq\varepsilon$ holds for all $\varepsilon\in(0,
\infty)$.
Therefore, we obtain that $\mathop{\prob\mbox{-}\lim}_{n \to\infty
}[L^n, L^n]_T = 0$, which completes
the proof.
\end{pf}

\subsection{\texorpdfstring{Preliminaries toward proving Theorem \protect\ref{thmutilmax}}
{Preliminaries toward proving Theorem 1.4}}
\label{subsecpreliminariesforproofofKS}

Consider a wealth-proc\-ess set $\X$. Define $\X^\circ$, the \textit
{process-polar} of $\X$, as the set of all nonnegative c\`adl\`ag
adapted processes $Y$ such that $Y_0 \leq1$ and $Y X$ is a $\prob
$-supermartingale for all \mbox{$X \in\X$}. Similarly, define $\X^{\circ
\circ}$, the
\textit{process-bipolar} of $\X$, as the set of all nonnegative
c\`adl\`ag adapted processes $X$ such that $X_0 \leq1 $ and $Y X$ is a
$\prob
$-supermartingale for all \mbox{$Y \in\X^\circ$}. (The terminology of the
process-polar and the process-bipolar was introduced in~\cite{MR1883202}.)

By definition, it is clear that $\X\subseteq\X^{\circ\circ}$ holds
for any
wealth-proce\-ss set $\X$---actually, one can provide a very concrete
description of the structure of $\X^{\circ\circ}$. Suppose that $\X
$ is a\vadjust{\goodbreak}
wealth-process set and that condition $\mathrm{NA}_1$ holds---in particular,
by Theorem~\ref{thmsmarts}, $\X\subseteq\Sem$. In~\cite{MR1883202},
and using the terminology of that paper, it is shown that $\X^{\circ
\circ}$ is the
smallest set of nonnegative c\`adl\`ag adapted processes that includes
$\X$ and is fork-convex, process-solid and Fatou-closed. The following
statement repeats this structural result for the process-bipolar, in a
slightly altered way to be useful later in the paper.\vspace*{-1pt}
%
\begin{theorem}[({\v Z}itkovi\'{c}~\cite{MR1883202})]
\label{thmprocess-bipolar}
Let $\X$ be a wealth-process set such that $\mathrm{NA}_1$ holds. Then,
$X \in\X^{\circ\circ}$ if and only if there exists an $\X$-valued sequence
$(X^n)_{\nin}$ and a sequence $(A^n)_{\nin}$ of nondecreasing adapted
c\`adl\`ag processes with $0 \leq A^n \leq1$ for each $\nin$ such that
$\mathsf{F}\mbox{-}\lim_{n \to\infty}X^n(1 - A^n) = X$.\vspace*{-1pt}
\end{theorem}

If follows from Theorem~\ref{thmprocess-bipolar} above that, if
condition $\mathrm{NA}_1$ is valid for a wealth-process set $\X$, the set
inclusions $\X\subseteq\oxs\subseteq\overline{\X}{}^{\mathsf
{F}}\subseteq\X^{\circ\circ}$ hold.

The following result regarding ``forward convex convergence'' will be
used twice in the sequel.\vspace*{-1pt}
%
\begin{lem} \label{lemfcc}
Let $\X$ be a wealth-process set such that $\mathrm{NA}_1$ holds.
Consider any $\X$-valued sequence $(X^n)_{\nin}$. Then, there exists an
$\X$-valued sequence $(\chi^n)_{\nin}$, with each $\chi^n$
belonging in
the convex hull of $ \{X^k \such k \geq n \}$, as well as
some $\chi
\in\overline{\X}{}^{\mathsf{F}}\subseteq\X^{\circ\circ}$ such
that $\mathsf{F}\mbox{-}\lim_{n \to\infty}\chi^n = \chi$.\vspace*{-1pt}
\end{lem}
\begin{pf}
In the notation of Theorem~\ref{thmmain}, consider the strictly
positive process $\hX\equiv\hX^\prob\in\oxs$ and define
$\widetilde
{\X} \dfn\{ X / \hX\such X \in\X\}$. It is straightforward
to check that $\widetilde{\X}$ is also a wealth-process set in the
sense of Definition~\ref{defnwealth-process}. All elements of
$\widetilde{\X}$ are nonnegative c\`adl\`ag $\prob$-supermartingales
starting from unit value. For the given $\X$-valued sequence
$(X^n)_{\nin}$, consider the $\widetilde{X}$-valued sequence
$(\widetilde{X}{}^n)_{\nin}$ defined via $\widetilde{X}{}^n \dfn X^n /
\hX$
for all $\nin$. Then,~\cite{MR1469917}, Lemma~5.2(1), implies that there
exists an $\X$-valued sequence $(\widetilde{\chi}{}^n)_{\nin}$, with each
$\widetilde{\chi}{}^n$ being in the convex hull of $ \{ \widetilde
{X}{}^k \such k \geq n \}$, as well as some nonnegative c\`adl\`ag
$\prob$-supermartingale $\widetilde{\chi}$ such that $\mathsf
{F}\mbox{-}\lim_{n \to\infty}\widetilde
{\chi}{}^n = \widetilde{\chi}$. Defining $\chi^n \dfn\hX\widetilde
{\chi
}{}^n$ for all $\nin$ and $\chi\dfn\hX\widetilde{\chi}$, the statement
of Lemma~\ref{lemfcc} immediately follows.\vspace*{-1pt}
\end{pf}

We pause for an interesting remark that will be soon useful. Assuming
condition $\mathrm{NA}_1$ on a wealth-process set $\X$, note that
$\hY\dfn1
/ \hX^\prob$ (in the notation of Theorem~\ref{thmmain}) is a strictly
positive process in $\X^\circ$---in fact, it is easy to show that the
converse also holds: existence of a strictly positive process in $\X
^\circ$
implies condition $\mathrm{NA}_1$.

Proposition~\ref{propbipolarduality} that follows (a static version
of ``bipolarity,'' a topic taken up in a general $\lzp$ setting in
\cite
{MR1768009}) is exactly the result that will allow us to use the
abstract formulation of results on expected utility maximization from~\cite{MR1722287} and~\cite{MR2023886}.\vspace*{-1pt}
%
\begin{prop} \label{propbipolarduality}
Suppose that $\X$ is a wealth-process set and that condition
$\mathrm{NA}_1$ is in force. Define\vadjust{\goodbreak} $\C\dfn\{X_T \such X \in
\X^{\circ\circ} \}$ and
$\D\dfn\{Y_T \such Y \in\X^\circ\}$. Then, we have
the following:
\begin{itemize}
\item for $g \in\lzp$, $g \in\C$ holds if and only if $\expec[h g]
\leq1$ holds for all $h \in\D$;
\item for $h \in\lzp$, $h \in\D$ holds if and only if $\expec[h g]
\leq1$ holds for all $g \in\C$.
\end{itemize}
\end{prop}
\begin{pf}
If $g \in\C$ and $h \in\D$, $\expec[hg] \leq1$ trivially holds.

Let $g \in\lzp$ be such that $\sup_{h \in\D}\expec[hg] \leq1$. We
shall show that there exists $X \in\X^{\circ\circ}$ such that $X_T
= g$. As
mentioned before the statement of Proposition~\ref{propbipolarduality},
under condition $\mathrm{NA}_1$ there exists a strictly positive
$\hY\in\X^\circ$; replacing, in obvious notation, $\X$ and $\X^{\circ
\circ}$ by
$\hY
\X$ and $\hY\X^{\circ\circ}$ and $\X^\circ$ by $(1/ \hY) \X^\circ$, we
may (and shall)
assume that $1 \in\X^\circ$. Let $\X^\circ_{++}$ be the set of all strictly
positive processes in $\X^\circ$. For all $t \in[0,T]$, define the (a
priori, possibly infinite-valued) $\F_t$-measurable random variable
$X^0_t \dfn\mathop{\operatorname{ess}\sup}_{Y \in\X^\circ_{++}}
\expec[(Y_T / Y_t) g \such\F_t ]$. As $\X^\circ_{++}$ is easily seen
to be fork-convex, the class of
random variables $ \{\expec[(Y_T / Y_t) g \such\F_t
] \such Y \in\X^\circ_{++} \}$ is upwards directed. (For the
definition of upwards
directed collections of random variables and their connection with the
notion of essential supremum, see~\cite{MR2169807}, Theorem A.32 in
Appendix A.5.) Furthermore, the fork-convexity of $\X^\circ_{++}$ combined
with the fact that $1 \in\X^\circ_{++}$ implies that $(Y_T / Y_t)
\in\D$
holds for all $Y \in\X^\circ_{++}$ and $t \in[0, T]$; therefore,
$\expec
[\expec[(Y_T / Y_t) g \such\F_t ] ] = \expec
[(Y_T / Y_t) g ] \leq1$ holds for all $Y \in\X^\circ_{++}$. It follows
that $\expec
[X_t^0] \leq1$ for all $t \in[0, T]$; in particular, $X^0_t \in\lzp$
for all $t \in[0, T]$. It is straightforward to check that $Y X^0$ is
a nonnegative supermartingale for all $Y \in\X^\circ_{++}$. In particular,
there exists a c\`adl\`ag process $X$ that coincides with the
right-continuous version of $X^0$ (for the terminal value, this means
$X_T = X^0_T = g$); then, the conditional version of Fatou's lemma
implies again that $Y X$ is a a nonnegative supermartingale for all $Y
\in\X^\circ_{++}$. For any fixed $Y \in\X^\circ$, $Y^n \dfn
(n^{-1} + (1 - n^{-1}) Y ) \in\X^\circ_{++}$ for all $\nin$.
Therefore, $Y^n X$ is a
supermartingale for all $\nin$; sending $n \to\infty$ and using the
conditional version of Fatou's lemma, we conclude that $Y X$ is a
supermartingale for all $Y \in\X^\circ$. Also, $X_0 \leq\liminf_{t
\downarrow0} \expec[X^0_t ] \leq1$. By the definition of the
process-bipolar, it follows that $X \in\X^{\circ\circ}$; since $X_T
= g$, we conclude.

In a completely similar way, it can be shown that if $h \in\lzp$ is
such that $\sup_{g \in\C}\expec[h g] \leq1$, then there exists $Y
\in
\Y$ such that $Y_T = h$. One needs to use the fork-convexity of $\X
^{\circ\circ}$
as well as the fact that $\X^\circ$ is the set of all c\`adl\`ag adapted
processes $Y$ with $Y_0 \leq1$ and such that $Y X$ is a nonnegative
supermartingale for all $X \in\X^{\circ\circ}$. Indeed, this last
fact follows
from the filtered bipolar theorem and Lemma 1 (with $\G= \F_0$) in
\cite{MR1883202}, since the process-bipolar of $\X^\circ$ coincides with
$\X^\circ
$ itself.
\end{pf}

\subsection{\texorpdfstring{Proof of Theorem \protect\ref{thmutilmax}}
{Proof of Theorem 1.4}}
\label{subsecproofofKS}

We retain all notation from Section
\ref{subsecpreliminariesforproofofKS}. In accordance with the
definition of $u$ from Section~\ref{subsecresults}, for $x \in(0,
\infty)$ define $\X^{\circ\circ}(x) \dfn\{x X \such X
\in\X^{\circ\circ} \}$ and $u^{\circ\circ}(x) = \sup_{X
\in\X^{\circ\circ}(x)} \expec[ U(X_T) ]$. The first thing to settle is
that the functions $u$ and $u^{\circ\circ}$ coincide.
%
\begin{lem} \label{lemequalvaluefunct}
Let $\X$ be a wealth process set with $1 \in\X$, such that
$\mathrm{NA}_1$ holds. Then, $u = u^{\circ\circ}$.
\end{lem}
\begin{pf}
Of course, $u \leq u^{\circ\circ}$ always holds; by way of
contradiction, assume
that $u(x) < u^{\circ\circ}(x)$ for some $x \in(0, \infty)$. Pick
$X \in\X^{\circ\circ}
(x)$ such that $\expec[ U(X_T) ] > u(x)$. Recalling Theorem
\ref{thmprocess-bipolar}, consider an $\X(x)$-valued sequence
$(X^n)_{\nin}$ and a sequence $(A^n)_{\nin}$ of nondecreasing adapted
c\`adl\`ag processes with $0 \leq A^n \leq1$ for each $\nin$ such that
$\mathsf{F}\mbox{-}\lim_{n \to\infty}X^n(1 - A^n) = X$. By Lemma
\ref{lemfcc}, there exists an $\X
(x)$-valued sequence $(\chi^n)_{\nin}$, with each $\chi^n$ being in the
convex hull of $ \{X^k \such k \geq n \}$, as well as some
$\chi\in
\X^{\circ\circ}(x)$ such that $\mathsf{F}\mbox{-}\lim_{n \to
\infty}\chi^n = \chi$. It is clear that $X_T \leq
\chi_T$ holds---therefore, $\expec[ U(\chi_T) ] \geq\expec
[
U(X_T) ] > u(x)$. It follows that we may (and shall) assume that
there exists $X \in\X^{\circ\circ}(x)$ such that $\expec[
U(X_T) ] >
u(x)$, as well as an $\X(x)$-valued sequence $(X^n)_{\nin}$ such that
$\mathsf{F}\mbox{-}\lim_{n \to\infty}X^n = X$. For $\kin$, define
the process $\tX^k \dfn(1/k)x + (1 -
1/k) X$; then $\tX^k \in\X^{\circ\circ}(x)$ for all $\kin$. Since
$\expec
[ 0
\wedge U(X_T) ] > - \infty$, the monotone convergence theorem
implies that there exists $K \in\Natural$ such that, with $\psi\dfn
\tX^K$, $\expec[ U(\psi_T) ] > u(x)$ holds. Now, for all
$\nin
$, define $\psi^n \dfn(1/K)x + (1 - 1/K) X^n$, so that $\psi^n \in
\X
(x)$. Note that $\mathsf{F}\mbox{-}\lim_{n \to\infty}\psi^n =
\psi$; in particular, $\prob[
\lim_{n \to\infty}
\psi^n_T = \psi_T ] = 1$. Since $\psi^n_T \geq x/K$ holds for all
$\nin$, using Fatou's lemma we obtain $\liminf_{n \to\infty} \expec
[U (\psi^n_T) ] \geq\expec[ U(\psi_T) ] >
u(x)$, which
contradicts the definition of $u$.
\end{pf}

According to Lemma~\ref{lemequalvaluefunct}, (\ref{eqFIN-DUAL})
holds with $u^{\circ\circ}$ replacing $u$ there. Fix $x \in(0,
\infty)$. In view
of Proposition~\ref{propbipolarduality}, under the assumptions of
Theorem~\ref{thmmain} one can use the abstract results of the utility
maximization problem in~\cite{MR1722287} and the results of \cite
{MR2023886} on the existence of the optimal wealth process, to show the
existence of a strictly positive $\hX\in\X^{\circ\circ}$ with $\hX_0 =
1$ such
that $\expec[ U(x \hX_T) ] = u^{\circ\circ}(x) = u(x) <
\infty$, as well
as the existence of a strictly positive $\hY\in\X^\circ$ such that
$\hY_0
= 1$ and $\hY\hX$ is a uniformly integrable martingale under $\prob$.
Define a probability $\qprob\sim\prob$ via $\ud\qprob= (\hY_T \hX_T) \ud
\prob$. Pick $X \in\X$, $t \in[0, T]$ and $s \in[0, t]$.
With ``$\expecq$'' denoting expectation under $\qprob$,
\[
\expecq\biggl[\frac{X_t}{\hX_t} \Bigm|\F_s \biggr] =
\frac
{1}{\hY_s
\hX_s} \expecp\biggl[{\hY_t \hX_t}
\frac{X_t}{\hX_t} \Bigm|\F_s \biggr] = \frac
{1}{\hY_s \hX_s} \expecp[
\hY_t X_t \such\F_s ] \leq
\frac
{1}{\hY_s
\hX_s} \hY_s X_s = \frac{X_s}{\hX_s},
\]
that is, $X / \hX$ is a $\qprob$-supermartingale. Using the conditional
version of Fatou's lemma, we can further deduce that $X / \hX$ is a
$\qprob$-supermartingale for all $X \in\overline{\X}{}^{\mathsf{F}}$.

We shall show now that $\hX\in\oxs$, which will establish both
statement (1) of Theorem~\ref{thmutilmax} and the part of statement
(2) of Theorem~\ref{thmutilmax} regarding existence of optimal
wealth processes. First, we show that $\hX\in\overline{\X}{}^{\mathsf
{F}}$. Since $\hX\in
\X^{\circ\circ}
$, in view of Theorem~\ref{thmprocess-bipolar} consider an $\X
$-valued sequence $(X^n)_{\nin}$ and a sequence of nondecreasing
adapted c\`adl\`ag processes with $0 \leq A^n \leq1$ for each $\nin$
such that $\mathsf{F}\mbox{-}\lim_{n \to\infty}X^n(1 - A^n) = \hX
$. By Lemma~\ref{lemfcc}, consider
an $\X$-valued sequence $(\chi^n)_{\nin}$, with each $\chi^n$ being in
the convex hull of $ \{X^k \such k \geq n \}$, as well as
some $\chi
\in\overline{\X}{}^{\mathsf{F}}$ such that $\mathsf{F}\mbox{-}\lim_{n \to
\infty}\chi^n = \chi$. From the two limiting
relationships, one can deduce that $\hX\leq\chi$. 
According to the preceding paragraph, $\chi/ \hX$ is a nonnegative
$\qprob$-supermartingale with $\chi_0 / \hX_0 \leq1$. This last fact
combined with $\chi/ \hX\geq1$ is easily seen to imply that $\chi=
\hX$---in other words, that $\hX\in\overline{\X}{}^{\mathsf{F}}$.
In order to actually show
that $\hX\in\oxs$, note that $ ( \chi^n / \hX)_{\nin}$
is a
sequence of nonnegative $\qprob$-supermartingales with $\chi^n_0 /
\hX_0 = 1$ and $\qprob[ \lim_{n \to\infty} ( \chi^n_T / \hX_T ) = 1
]
= 1$. Recalling the arguments\vspace*{1pt} of Section~\ref{subsecproofofthmmain},
we deduce that $\mathcal{S}\mbox{-}\lim_{n \to\infty}\chi^n = \hX
$, which implies that $\hX
\in\oxs$.

We now move to establish the existence of an approximating sequence as
required in statement (2) of Theorem~\ref{thmutilmax}, which will
complete the proof. Fix $x \in(0, \infty)$. Let\vspace*{1pt} $\hX\equiv\hX(x)$ be
the optimizer in $\oxs(x)$ of the utility maximization problem. We know
that there exists an $\X(x)$-valued\vspace*{1pt} sequence $(\tX^k)_{\kin}$ such that
$\mathcal{S}\mbox{-}\lim_{k \to\infty}\tX^k = \hX$. However, it
might not hold that $\lim_{k \to\infty}
\expec[U(\tX^k_T)] = u(x)$. To circumvent this issue, set $\hX^n
\dfn
(1/n) x + (1 - 1/n) \hX$ for $\nin$. Note\vspace*{1pt} that $\mathcal{S}\mbox
{-}\lim_{n \to\infty}\hX^n = \hX$ and
$\lim_{n \to\infty}\expec[U(\hX^n_T)] = \expec[U(\hX_T)]$ hold.
For each $\nin$,
pick $k_n \in\Natural$ such that, with $X^n \dfn(1/n) x + (1 - 1/n)
\tX^{k_n}$, we have $ \lceil X^n - \hX^n \rceil_{\Sem}
\leq n^{-1}$ and
$\expec[U(X^n_T)] \geq\expec[U(\hX^n_T)] - n^{-1}$, the latter being
feasible in view of Fatou's lemma. As $\mathcal{S}\mbox{-}\lim_{n
\to\infty}\hX^n = \hX$ and $\lim_{n \to\infty}
\expec[U(\hX^n_T)] = \expec[U(\hX_T)]$, we conclude that the sequence
$(X^n)_{\nin}$ satisfies the requirements of statement (2) of Theorem
\ref{thmutilmax}.

\section*{Acknowledgments}

The author would like to thank two anonymous referees and the Associate
Editor who dealt with the paper for valuable input and constructive
comments.



\printaddresses


\begin{thebibliography}{28}

\bibitem{MR1849424}
\begin{barticle}[mr]
\bauthor{\bsnm{Becherer},~\bfnm{Dirk}\binits{D.}}
(\byear{2001}).
\btitle{The numeraire portfolio for unbounded semimartingales}.
\bjournal{Finance Stoch.}
\bvolume{5}
\bpages{327--341}.
\bid{doi={10.1007/PL00013535}, issn={0949-2984}, mr={1849424}}
\bptok{imsref}%
\end{barticle}
\endbibitem

\bibitem{BSV}
\begin{barticle}[mr]
\bauthor{\bsnm{Beiglb{\"o}ck},~\bfnm{Mathias}\binits{M.}},
  \bauthor{\bsnm{Schachermayer},~\bfnm{Walter}\binits{W.}} \AND
  \bauthor{\bsnm{Veliyev},~\bfnm{Bezirgen}\binits{B.}}
(\byear{2011}).
\btitle{A direct proof of the {B}ichteler--{D}ellacherie theorem and connections
  to arbitrage}.
\bjournal{Ann. Probab.}
\bvolume{39}
\bpages{2424--2440}.
\bid{doi={10.1214/10-AOP602}, issn={0091-1798}, mr={2932672}}
\bptnote{check year}%
\bptok{imsref}%
\end{barticle}
\endbibitem

\bibitem{BDMKR}
\begin{barticle}[auto:STB|2012/12/05|11:57:16]
\bauthor{\bsnm{Bj{\"o}rk},~\bfnm{T.}\binits{T.}},
  \bauthor{\bsnm{Di~Masi},~\bfnm{G.}\binits{G.}},
  \bauthor{\bsnm{Kabanov},~\bfnm{Y.}\binits{Y.}} \AND
  \bauthor{\bsnm{Runggaldier},~\bfnm{W.}\binits{W.}}
(\byear{1997}).
\btitle{Towards a general theory of bond markets}.
\bjournal{Finance Stoch.}
\bvolume{1}
\bpages{141--174}.
\bid{mr={2976683}}
\bptok{imsref}%
\end{barticle}
\endbibitem

\bibitem{MR1768009}
\begin{bincollection}[mr]
\bauthor{\bsnm{Brannath},~\bfnm{W.}\binits{W.}} \AND
  \bauthor{\bsnm{Schachermayer},~\bfnm{W.}\binits{W.}}
(\byear{1999}).
\btitle{A bipolar theorem for $L^0_+ (\Omega ,\mathcal{F}, \mathbb{P})$}.
In \bbooktitle{S\'eminaire de {P}robabilit\'es, {XXXIII}}.
\bseries{Lecture Notes in Math.}
\bvolume{1709}
\bpages{349--354}.
\bpublisher{Springer}, \blocation{Berlin}.
\bid{doi={10.1007/BFb0096525}, mr={1768009}}
\bptok{imsref}%
\end{bincollection}
\endbibitem

\bibitem{Cz-Sch}
\begin{bincollection}[mr]
\bauthor{\bsnm{Czichowsky},~\bfnm{Christoph}\binits{C.}} \AND
  \bauthor{\bsnm{Schweizer},~\bfnm{Martin}\binits{M.}}
(\byear{2011}).
\btitle{Closedness in the semimartingale topology for spaces of stochastic
  integrals with constrained integrands}.
In \bbooktitle{S\'eminaire de {P}robabilit\'es {XLIII}}.
\bseries{Lecture Notes in Math.}
\bvolume{2006}
\bpages{413--436}.
\bpublisher{Springer}, \blocation{Berlin}.
\bid{doi={10.1007/978-3-642-15217-7_18}, mr={2790384}}
\bptok{imsref}%
\end{bincollection}
\endbibitem

\bibitem{MR2178505}
\begin{barticle}[mr]
\bauthor{\bsnm{De~Donno},~\bfnm{M.}\binits{M.}},
  \bauthor{\bsnm{Guasoni},~\bfnm{P.}\binits{P.}} \AND
  \bauthor{\bsnm{Pratelli},~\bfnm{M.}\binits{M.}}
(\byear{2005}).
\btitle{Super-replication and utility maximization in large financial markets}.
\bjournal{Stochastic Process. Appl.}
\bvolume{115}
\bpages{2006--2022}.
\bid{doi={10.1016/j.spa.2005.06.010}, issn={0304-4149}, mr={2178505}}
\bptok{imsref}%
\end{barticle}
\endbibitem

\bibitem{MR2187311}
\begin{barticle}[mr]
\bauthor{\bsnm{De~Donno},~\bfnm{M.}\binits{M.}} \AND
  \bauthor{\bsnm{Pratelli},~\bfnm{M.}\binits{M.}}
(\byear{2005}).
\btitle{A theory of stochastic integration for bond markets}.
\bjournal{Ann. Appl. Probab.}
\bvolume{15}
\bpages{2773--2791}.
\bid{doi={10.1214/105051605000000548}, issn={1050-5164}, mr={2187311}}
\bptok{imsref}%
\end{barticle}
\endbibitem

\bibitem{MR1304434}
\begin{barticle}[mr]
\bauthor{\bsnm{Delbaen},~\bfnm{Freddy}\binits{F.}} \AND
  \bauthor{\bsnm{Schachermayer},~\bfnm{Walter}\binits{W.}}
(\byear{1994}).
\btitle{A general version of the fundamental theorem of asset pricing}.
\bjournal{Math. Ann.}
\bvolume{300}
\bpages{463--520}.
\bid{doi={10.1007/BF01450498}, issn={0025-5831}, mr={1304434}}
\bptok{imsref}%
\end{barticle}
\endbibitem

\bibitem{MR1381678}
\begin{barticle}[mr]
\bauthor{\bsnm{Delbaen},~\bfnm{Freddy}\binits{F.}} \AND
  \bauthor{\bsnm{Schachermayer},~\bfnm{Walter}\binits{W.}}
(\byear{1995}).
\btitle{The no-arbitrage property under a change of num\'eraire}.
\bjournal{Stochastics Stochastics Rep.}
\bvolume{53}
\bpages{213--226}.
\bid{issn={1045-1129}, mr={1381678}}
\bptok{imsref}%
\end{barticle}
\endbibitem

\bibitem{MR2722836}
\begin{bbook}[mr]
\bauthor{\bsnm{Durrett},~\bfnm{Rick}\binits{R.}}
(\byear{2010}).
\btitle{Probability: Theory and Examples},
\bedition{4th} ed.
\bpublisher{Cambridge Univ. Press}, \blocation{Cambridge}.
\bid{mr={2722836}}
\bptok{imsref}%
\end{bbook}
\endbibitem

\bibitem{MR544800}
\begin{bincollection}[mr]
\bauthor{\bsnm{Emery},~\bfnm{M.}\binits{M.}}
(\byear{1979}).
\btitle{Une topologie sur l'espace des semimartingales}.
In \bbooktitle{S\'eminaire de {P}robabilit\'es, {XIII} ({U}niv. {S}trasbourg,
  {S}trasbourg, 1977/78)}.
\bseries{Lecture Notes in Math.}
\bvolume{721}
\bpages{260--280}.
\bpublisher{Springer}, \blocation{Berlin}.
\bid{mr={0544800}}
\bptok{imsref}%
\end{bincollection}
\endbibitem

\bibitem{MR1469917}
\begin{barticle}[mr]
\bauthor{\bsnm{F{\"o}llmer},~\bfnm{H.}\binits{H.}} \AND
  \bauthor{\bsnm{Kramkov},~\bfnm{D.}\binits{D.}}
(\byear{1997}).
\btitle{Optional decompositions under constraints}.
\bjournal{Probab. Theory Related Fields}
\bvolume{109}
\bpages{1--25}.
\bid{doi={10.1007/s004400050122}, issn={0178-8051}, mr={1469917}}
\bptok{imsref}%
\end{barticle}
\endbibitem

\bibitem{MR2169807}
\begin{bbook}[mr]
\bauthor{\bsnm{F{\"o}llmer},~\bfnm{Hans}\binits{H.}} \AND
  \bauthor{\bsnm{Schied},~\bfnm{Alexander}\binits{A.}}
(\byear{2004}).
\btitle{Stochastic Finance: An Introduction in Discrete Time},
\bedition{extended} ed.
\bseries{de Gruyter Studies in Mathematics}
\bvolume{27}.
\bpublisher{de Gruyter}, \blocation{Berlin}.
\bid{doi={10.1515/9783110212075}, mr={2169807}}
\bptok{imsref}%
\end{bbook}
\endbibitem

\bibitem{MR2335830}
\begin{barticle}[mr]
\bauthor{\bsnm{Karatzas},~\bfnm{Ioannis}\binits{I.}} \AND
  \bauthor{\bsnm{Kardaras},~\bfnm{Constantinos}\binits{C.}}
(\byear{2007}).
\btitle{The num\'eraire portfolio in semimartingale financial models}.
\bjournal{Finance Stoch.}
\bvolume{11}
\bpages{447--493}.
\bid{doi={10.1007/s00780-007-0047-3}, issn={0949-2984}, mr={2335830}}
\bptok{imsref}%
\end{barticle}
\endbibitem

\bibitem{Kar10}
\begin{barticle}[auto:STB|2012/12/05|11:57:16]
\bauthor{\bsnm{Kardaras},~\bfnm{C.}\binits{C.}}
(\byear{2013}).
\btitle{Generalized supermartingale deflators under limited information}.
\bjournal{Math. Finance}
\bvolume{23}
\bpages{186--197}.
\bptok{imsref}%
\end{barticle}
\endbibitem

\bibitem{Kar11}
\begin{barticle}[auto:STB|2012/12/05|11:57:16]
\bauthor{\bsnm{Kardaras},~\bfnm{C.}\binits{C.}}
(\byear{2012}).
\btitle{Market viability via absence of arbitrage of the first kind}.
\bjournal{Finance Stoch.}
\bvolume{16}
\bpages{651--667}.
\bptok{imsref}%
\end{barticle}
\endbibitem

\bibitem{Kar11b}
\begin{barticle}[mr]
\bauthor{\bsnm{Kardaras},~\bfnm{Constantinos}\binits{C.}}
(\byear{2012}).
\btitle{A structural characterization of num\'eraires of convex sets of
  nonnegative random variables}.
\bjournal{Positivity}
\bvolume{16}
\bpages{245--253}.
\bid{doi={10.1007/s11117-011-0120-1}, issn={1385-1292}, mr={2929089}}
\bptnote{check year}%
\bptok{imsref}%
\end{barticle}
\endbibitem

\bibitem{MR2832419}
\begin{barticle}[mr]
\bauthor{\bsnm{Kardaras},~\bfnm{Constantinos}\binits{C.}} \AND
  \bauthor{\bsnm{Platen},~\bfnm{Eckhard}\binits{E.}}
(\byear{2011}).
\btitle{On the semimartingale property of discounted asset-price processes}.
\bjournal{Stochastic Process. Appl.}
\bvolume{121}
\bpages{2678--2691}.
\bid{doi={10.1016/j.spa.2011.06.010}, issn={0304-4149}, mr={2832419}}
\bptok{imsref}%
\end{barticle}
\endbibitem

\bibitem{MR1722287}
\begin{barticle}[mr]
\bauthor{\bsnm{Kramkov},~\bfnm{D.}\binits{D.}} \AND
  \bauthor{\bsnm{Schachermayer},~\bfnm{W.}\binits{W.}}
(\byear{1999}).
\btitle{The asymptotic elasticity of utility functions and optimal investment
  in incomplete markets}.
\bjournal{Ann. Appl. Probab.}
\bvolume{9}
\bpages{904--950}.
\bid{doi={10.1214/aoap/1029962818}, issn={1050-5164}, mr={1722287}}
\bptok{imsref}%
\end{barticle}
\endbibitem

\bibitem{MR2023886}
\begin{barticle}[mr]
\bauthor{\bsnm{Kramkov},~\bfnm{D.}\binits{D.}} \AND
  \bauthor{\bsnm{Schachermayer},~\bfnm{W.}\binits{W.}}
(\byear{2003}).
\btitle{Necessary and sufficient conditions in the problem of optimal
  investment in incomplete markets}.
\bjournal{Ann. Appl. Probab.}
\bvolume{13}
\bpages{1504--1516}.
\bid{doi={10.1214/aoap/1069786508}, issn={1050-5164}, mr={2023886}}
\bptok{imsref}%
\end{barticle}
\endbibitem

\bibitem{MR2260066}
\begin{barticle}[mr]
\bauthor{\bsnm{Kramkov},~\bfnm{Dmitry}\binits{D.}} \AND
  \bauthor{\bsnm{S{\^{\i}}rbu},~\bfnm{Mihai}\binits{M.}}
(\byear{2006}).
\btitle{On the two-times differentiability of the value functions in the
  problem of optimal investment in incomplete markets}.
\bjournal{Ann. Appl. Probab.}
\bvolume{16}
\bpages{1352--1384}.
\bid{doi={10.1214/105051606000000259}, issn={1050-5164}, mr={2260066}}
\bptok{imsref}%
\end{barticle}
\endbibitem

\bibitem{MR2288717}
\begin{barticle}[mr]
\bauthor{\bsnm{Kramkov},~\bfnm{Dmitry}\binits{D.}} \AND
  \bauthor{\bsnm{S{\^{\i}}rbu},~\bfnm{Mihai}\binits{M.}}
(\byear{2006}).
\btitle{Sensitivity analysis of utility-based prices and risk-tolerance wealth
  processes}.
\bjournal{Ann. Appl. Probab.}
\bvolume{16}
\bpages{2140--2194}.
\bid{doi={10.1214/105051606000000529}, issn={1050-5164}, mr={2288717}}
\bptok{imsref}%
\end{barticle}
\endbibitem

\bibitem{Long90}
\begin{barticle}[auto:STB|2012/12/05|11:57:16]
\bauthor{\bsnm{Long},~\bfnm{J.~B.~J.}\binits{J.~B.~J.}}
(\byear{1990}).
\btitle{The num\'eraire portfolio}.
\bjournal{Journal of Financial Economics}
\bvolume{26}
\bpages{29--69}.
\bptok{imsref}%
\end{barticle}
\endbibitem

\bibitem{MR568256}
\begin{barticle}[mr]
\bauthor{\bsnm{M{\'e}min},~\bfnm{Jean}\binits{J.}}
(\byear{1980}).
\btitle{Espaces de semi martingales et changement de probabilit\'e}.
\bjournal{Z.~Wahrsch. Verw. Gebiete}
\bvolume{52}
\bpages{9--39}.
\bid{doi={10.1007/BF00534184}, issn={0044-3719}, mr={0568256}}
\bptok{imsref}%
\end{barticle}
\endbibitem

\bibitem{MR2273672}
\begin{bbook}[mr]
\bauthor{\bsnm{Protter},~\bfnm{Philip~E.}\binits{P.~E.}}
(\byear{2005}).
\btitle{Stochastic Integration and Differential Equations},
\bedition{2nd} ed.
\bseries{Stochastic Modelling and Applied Probability}
\bvolume{21}.
\bpublisher{Springer}, \blocation{Berlin}.
\bnote{Version 2.1, Corrected third printing}.
\bid{mr={2273672}}
\bptok{imsref}%
\end{bbook}
\endbibitem

\bibitem{Taka10}
\begin{bmisc}[auto:STB|2012/12/05|11:57:16]
\bauthor{\bsnm{Takaoka},~\bfnm{K.}\binits{K.}}
(\byear{2012}).
\bhowpublished{On the condition of no unbounded profit with bounded risk.
  \textit{Finance Stoch.} To appear. Available at
  \texttt{%
  \href{http://hermes-ir.lib.hit-u.ac.jp/rs/handle/10086/18812}{http://hermes-ir.lib.hit-u.ac.jp/rs/}
  \href{http://hermes-ir.lib.hit-u.ac.jp/rs/handle/10086/18812}{handle/10086/18812}}}.
\bptok{imsref}%
\end{bmisc}
\endbibitem

\bibitem{MR1883202}
\begin{barticle}[mr]
\bauthor{\bsnm{{\v{Z}}itkovi{\'c}},~\bfnm{Gordan}\binits{G.}}
(\byear{2002}).
\btitle{A filtered version of the bipolar theorem of {B}rannath and
  {S}chachermayer}.
\bjournal{J. Theoret. Probab.}
\bvolume{15}
\bpages{41--61}.
\bid{doi={10.1023/A:1013885121598}, issn={0894-9840}, mr={1883202}}
\bptok{imsref}%
\end{barticle}
\endbibitem

\end{thebibliography}
\end{document}